\documentclass[nonacm, sigconf]{acmart}

\makeatletter                   
\def\mdseries@tt{m}             
\makeatother                    

\usepackage[plain]{fancyref}
\usepackage[draft=true]{minted} 
\usepackage{color}
\usepackage{listings}
\usepackage{minted}
\usepackage{amsmath}
\usepackage{booktabs}
\usepackage{graphicx}
\usepackage{float}
\usepackage{url}
\floatstyle{boxed} 
\usepackage{subcaption}
\usepackage[toc,page]{appendix}

\usepackage{multirow}
\usepackage{mathtools}
\usepackage{mathrsfs}
\usepackage{calc}
\usepackage[neverdecrease]{paralist}

\usepackage{enumitem}

\newcommand{\para}[1]{\paragraph{#1}}

\DeclarePairedDelimiter{\ceil}{\lceil}{\rceil}
\DeclarePairedDelimiter\floor{\lfloor}{\rfloor}

\begin{document}
\sloppy                         

\title{Approximate Nearest Neighbour Search on Privacy-aware Encoding of User Locations to Identify Susceptible Infections in Simulated Epidemics}

\author{Chandan Biswas}
\affiliation{%
  \institution{Indian Statistical Institute, Kolkata}
}
\email{chandanbiswas08_r@isical.ac.in}

\author{Debasis Ganguly}
\affiliation{%
  \institution{IBM Research, Dublin, Ireland}
}
\email{debasis.ganguly1@ie.ibm.com}

\author{Ujjwal Bhattacharya}
\affiliation{%
  \institution{Indian Statistical Institute, Kolkata}
}
\email{ujjwal@isical.ac.in}


\begin{abstract}
Amidst an increasing number of infected cases during the Covid-19 pandemic, it is essential to trace, as early as possible, the susceptible people who might have been infected by the disease due to their close proximity with people who were tested positive for the virus. This early contact tracing is likely to limit the rate of spread of the infection within a locality.
In this paper, we investigate how effectively and efficiently can such a list of susceptible people be found given a list of infected persons and their locations. To address this problem from an information retrieval (search) perspective, we represent the location of each person at each time instant as a point in a vector space. By using the locations of the given list of infected persons as queries, we investigate the feasibility of applying \emph{approximate nearest neighbour} (ANN) based indexing and retrieval approaches to obtain a list of top-$k$ suspected users in real-time.
Since leveraging information from true user location data can lead to security and privacy concerns, we also investigate what effects does distance-preserving encoding methods have on the effectiveness of the ANN methods.
Experiments conducted on real and synthetic datasets demonstrate that the top-$k$ retrieved lists of susceptible users retrieved with existing ANN approaches (KD-tree and HNSW)
yield satisfactory precision and recall values, thus indicating that ANN approaches can potentially be applied in practice to facilitate real-time contact tracing even under the presence of imposed privacy constraints.

\end{abstract}


\maketitle

\section{Introduction}
The currently ongoing Covid-19 pandemic has spread at a rapidly accelerating rate since its inception. Standard epidemics analysis models, e.g., the SIR model \citep{sirmodel}, have stressed on the importance of finding the susceptible cases to flatten the growth rate of the spread of infection as early as possible. In this modern era of ubiquitous digital connectivity through mobile devices, a possible source of information for contact tracing is  the log of location traces in the form of GPS coordinates.

Since procuring such data for the purpose of contact tracing and using it in a restricted way (possibly by government organizations) is difficult and time-consuming due to the very sensitive nature of the data, a strong case needs to be made that how could such data be useful for controlling the spread of a pandemic.
The aim of this article is to to demonstrate a \emph{proof-of-the-concept} that with the availability of massive amounts of trajectory data, it is \emph{feasible} to develop a \emph{scalable system} that is both effective (in terms of identifying people susceptible to an infectious disease) and efficient (in terms of the time taken to identify the susceptible cases). We believe that this \emph{proof-of-the-concept} will encourage sharing (with restricted use) of such sensitive data in order to help mitigate epidemic situations.

In this paper, we formalize \emph{contact tracing} as a search problem within an Euclidean vector space. More concretely, each the \emph{state} of each person is represented as a point in a vector space, specifically of $4$ dimensions constituting $3$ dimensions for space ($3$ Cartesian coordinates corresponding to the spherical coordinates for latitude and longitude on the Earth's surface) and $1$ for time.
A given set of persons (those diagnosed as positive with the disease) then define the query points in this vector space.
People who were \emph{close} to these infected persons, in terms of both space and time (i.e. they were in approximately the same place at nearly the same time), also carry the risk of being infected with the disease. The objective then is to obtain a list of such susceptible people in real-time.
Figure \ref{fig:space-time} schematically depicts the idea.

The number of points represented in this vector space can rapidly grow in situations where either the geographic area represented is too large or too dense to start with, or the location traces need to be represented over a large duration of time (e.g. over several months). An exhaustive search for finding susceptible infection cases in this space is likely not to be feasible in terms of computation time. However, this formulation makes provision to investigate the use of approximate nearest neighbor (ANN) approaches, such as KD-trees \citep{silpa2008optimised}, and evaluate the effectiveness of such approximate approaches mainly in terms of relative recall with respect to the exhaustive search (i.e. how many such truly susceptible cases can the approximate algorithm find out). Ideally speaking, we could consider an ANN algorithm to be working well in this situation of contact tracing if it achieves a fair trade-off between the computation time and the recall relative to the exhaustive search (minimizing the former and maximizing the latter).


\para{Our Contributions}

The novelty of our work lies in investigating, under a laboratory based reproducible environment, the feasibility of ANN algorithms for contact tracing during epidemics. We, to the best of our knowledge, are not aware of any other work along this direction.
In particular, we conduct extensive experiments on a relatively large database ($24$M) of real GPS locations, and an even larger collection ($150$M) of synthetic data comprising random walks of simulated agents. The workflow of our experiments involves indexing a large collection of trajectory records, followed by simulating a number of records from this index as infected (representing the real-life situation of new cases of reported infection). 
Given the location trace of each infected person, we then find out a candidate list of persons and evaluate the retrieval effectiveness.
Additionally, since sharing true location data of real users across different organizations can potentially cause privacy concerns, we also investigate the feasibility of encoding the true locations with a distance-preserving linear transformation, e.g. \citep{ji2012super}. While such encoding has been shown to preserve privacy of data \citep{BiswasGRB19}, we investigate what effects can such an encoding have on the effectiveness of the ANN retrieval algorithms.

The findings of our experiments indicate that ANN based approaches do yield satisfactory recall even on encoded data. In terms of run-time, the ANN based approach achieves up to $17,000\times$ speed-up.
We emphasize that the scope of this paper is not to explore a novel ANN method but rather to study the feasibility of applying ANN methods for contact tracing in an epidemic situation.

\section{Background}
\textbf{Approximate nearest neighbors search.}
Existing studies in nearest neighbors (NN) search attempt to find the closest $k$ objects to a query point $q$ from a dataset $D$.
The KD-tree algorithm, as proposed in the classic paper \citep{bentley1975multidimensional}, is one of the most popular \emph{exact} nearest neighbors (NN) searching algorithm. Although it yields good results in low dimensional spaces, its effectiveness in terms of computation time and memory usage, tends to decrease for high dimensional spaces \citep{chavez2001searching}.
Since exact NN finding for high dimensional spaces takes a substantial amount of time, 
exact nearest neighbor (NN) search algorithms (e.g. the classic KD-tree) being computationally expensive, are rather intractable for large collections of embedded data in high dimensional spaces, thus leading towards research on approximate NN retrieval.
Approximate nearest neighbor (ANN) search finds applications
in content-based image retrieval. With the advent of deep learning based methods which represent image and text data in a joint embedding space of reals \citep{Frome13}, finding nearest neighbors in the data can be useful for various applications, such as image captioning \citep{Karpathy:2017}, `imagification' of documents \citep{Agrawal:2011} etc.

Generally speaking, existing ANN approaches can broadly be divided into the following categories.
Firstly, some approaches are memory-based relying on efficient data structures to compute only a limited number of exact distances \citep{CiacciaP00}.
Variations of the KD-tree data-structure to support approximate NN (ANN) also fall in this category \citep{arya1998optimal, beis1997shape, silpa2008optimised}.

Secondly, some approaches are hash-based which aim to design effective hash functions to preserve the spatial proximity of the points, i.e. map close points to the identical hash values \citep{andoni2006near,Andoni:2008,Bawa:2005}.
Locality sensitive hashing (LSH) \citep{andoni2006near} is the most popular hashing-based ANN search method which uses a number of different distance preserving (also called semantic) hash functions. Generally speaking, the effectiveness of these hashing-based techniques solely depends on the hash functions used. A significant volume of research has been directed towards improving the quality of these hash functions, such as the Super-bit LSH \citep{ji2012super}, kernel-LSH \citep{kulis2009kernelized}, randomized hashing with metric learning \citep{jain2008fast} etc.

The third category of approaches map data points to compact binary codes to reduce in-memory space and achieve fast exhaustive search
in Hamming space \citep{YunchaoGong:2011}. Product quantization (PQ) \citep{Jegou:2011} is a specific type of non-binary discrete encoding method used for either exhaustive search or non-exhaustive search with the help of inverted indexing. Supervised methods have been proposed to better fit the
quantization parameters to the underlying data distribution \citep{opq,KalantidisA14}.

The fourth category of approaches is based on metric inversion (MI), i.e. relying on pre-computing distances from a set of reference points (different from the data points). These distances are stored in the postings list corresponding to each reference term \citep{Amato:2014,ChavezGonzalez:2008,Chavez:2015}.
Among more successful approaches allowing
provision for an inverted index based secondary storage organization (with query driven dynamic loading of content in the primary memory) are the graph-based approaches - NSW and HNSW.
Navigable Small World (NSW) \citep{boguna2009navigability} is a graph with logarithmic or poly-logarithmic scaling of greedy graph routing \citep{boguna2009navigability}. \citet{malkov2014approximate} further improved NSW-based ANN search with a controlled hierarchy based approach, known as the Hierarchical NSW (HNSW).


\textbf{Trajectory search.}
Recent advancement in GPS technology has enabled led to everyday recording and storing large amounts of trajectory data of moving objects. This high volume trajectory can be very useful for trip recommendation \citep{shang2012user}, travel time and travel path optimization \citep{wang2014travel}, identifying driver expertise \citep{sun2018discovering} etc. There also exist a number of recent studies that about trajectory search given a particular query location \citep{chen2010searching, tang2011retrieving}, region of interest \citep{shang2017searching} or traveler's preference and activity \citep{shang2012user, shang2014personalized, zheng2013towards, wang2017answering}.

\subsection{Review of KD-Tree and HSNW}

Since our experiments are conducted with an in-memory and an indexing based ANN approach, specifically the KD-tree and HSNW, respectively, in this section we briefly review these two approaches. 

\textbf{KD-Tree.}
KD-tree is a multi-level space partitioning binary search tree data structure, where $K$ is the dimensionality of the search space.
Each node in the tree consists of $K$ keys (which comprise the data vector) and two pointers which points to the left sub-tree and right sub-trees.
The general idea in Kd-tree is to partition a given collection of points by hyperplanes perpendicular to the axes. Associated with each node is an integer $j$ ($0\leq j <K$) called the \emph{discriminator}, the role of which is to determine the direction (left or right) of a data point with respect to the splitting hyperplane (the hyperplane perpendicular to $j^{th}$ dimension's axis). The root node has the discriminator value $0$.
%
Insertion and searching in a KD-tree recursively traverses the tree determining the discriminator values at each level by computing the median of the values corresponding to the $j$-th dimension.


\textbf{HSNW.}
In general, proximity graph based methods constructs an index by preserving the links to closest neighbours for each individual data point. The basic greedy search algorithm on this proximity graph is very expensive due to curse of dimensionality and it gives relatively poor performance on data with well-separable clusters \citep{malkov2014approximate}. To address this limitation, \citet{malkov2014approximate} proposed navigable small world (NSW) graph based algorithm for solving the approximate nearest neighbour search problem. The NSW graph, $G(V,E)$, is a network with logarithmic or poly-logarithmic scalability of the greedy search algorithm \citep{kleinberg2000small}, where there is a one-one mapping between the vertex set $V$ of $G$ and the elements of the input dataset $X \subset \mathbb{R}^d$, the set of edges $E$ representing the link among the elements being determined by the following construction algorithm. The edge construction algorithm repeatedly connects a randomly selected node (a data point) with its nearest neighbor. More formally, $(u,v) \in E$ if $\vec{x}_u \in N(\vec{x}_v)$ or $\vec{x}_v \in N(\vec{x}_u)$, where $N(\vec{x})$ represents the neighbourhood of a point $\vec{x}$ in $X$.

The NSW algorithm was further improved by the Hierarchical NSW (HNSW) algorithm proposed in \citet{malkov2018efficient}.
The key idea of index construction and search strategy in HNSW is
to extend the graph structure of NSW into a hierarchy of a multi-layered structure, having the links separated by their characteristic distance scales.
The HSNW graph is constructed by consecutively inserting a node for each data point, where for each node inserted an integer $l=\floor{-\ln{(\mathcal{U}(0,1))}m_L}$ is chosen to determine the maximum level of the element, where $m_L$ is a normalization factor for level generation. The insertion procedure has two phases.
In the first phase a greedy algorithm starts from the top layer to find $e$ closest neighbours of the inserted element ($e$ is a parameter to control the search quality, its value in the first phase being set to $1$). In the second phase, search continues to the lower layer considering the closest neighbours found in first phase as entry points and the process repeats. 
The HSNW ANN search procedure is identical to the insertion algorithm for an element with layer $l=0$. The search result constitutes the closest neighbors found at the bottom-most layer.



\section{ANN Workflow}
\label{sec:method}

\subsection{Representation of Location Data}
The geo-locations of users (which in real life can be obtained from GPS locations of smart phones) are represented by `3'-dimensional points (\emph{2 space dimensions} corresponding to the location on the Earth's surface latitude, longitude and \emph{a time dimension} measured in system epochs). The path traced in this 3 dimensional space-time corresponds to the activity phase of a single user.

\begin{figure}[t]
    \centering
    \includegraphics[width = 0.8\columnwidth]{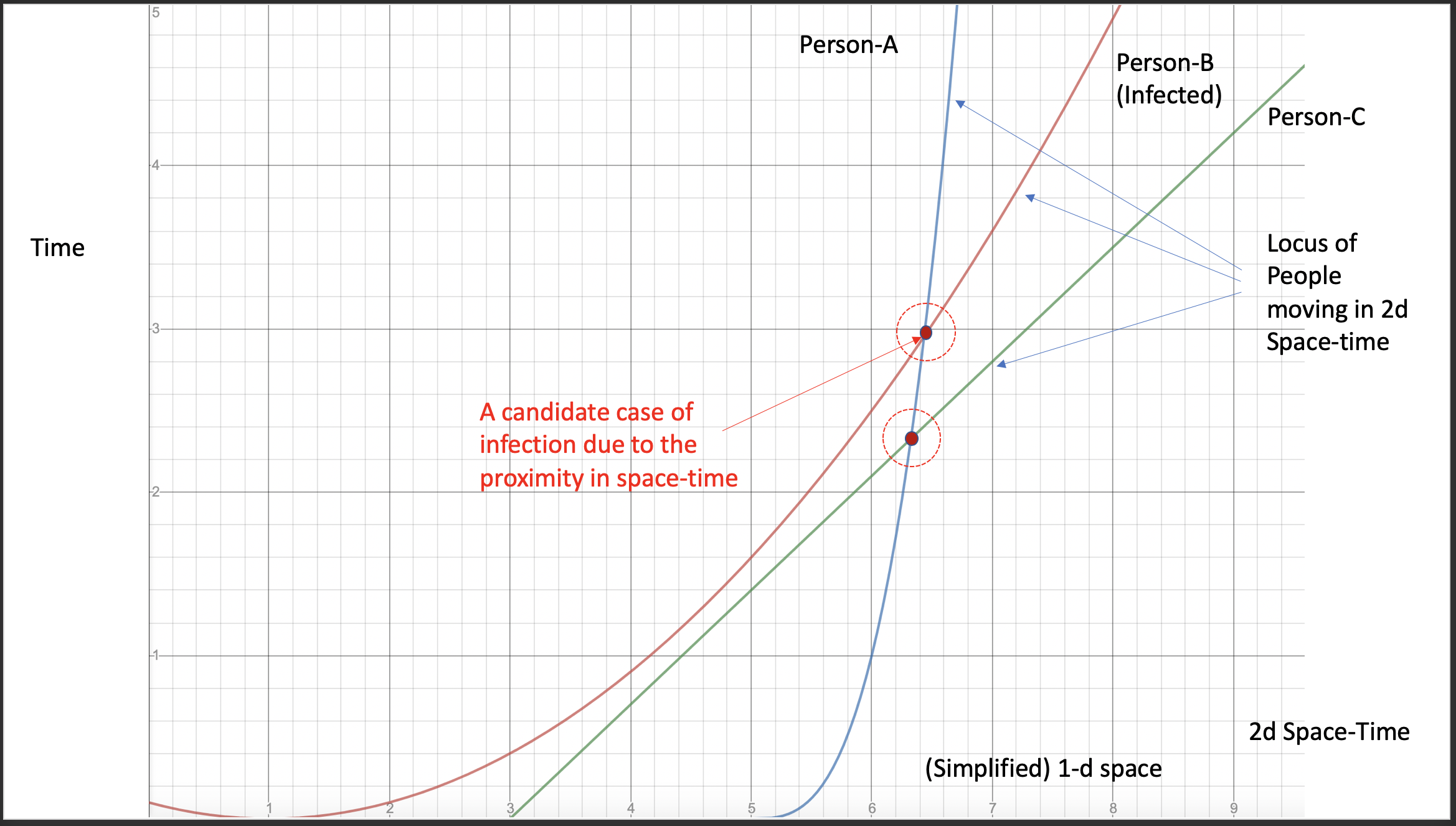}
    \caption{A simple visualization of a 2d space-time world.}
    \label{fig:space-time}
\end{figure}

Figure \ref{fig:space-time} shows a schematic visualization of a 2d space-time world. Each person is shown as a path (curve) in this space-time, i.e. each person is shown as a locus of changing positions (x coordinate) with respect to time. Values along the time dimension (y-axis in the figure) monotonically increase along the the x coordinate, or in other words the curves never loop down. Figure \ref{fig:space-time} shows two intersections of these locus curves. One of these is an intersection of a healthy person with an infected one (leaving the healthy person at a high risk of infection). The objective of the ANN based search is to \emph{automatically find all such possible intersections} given a large collection of each individual's location traces (curves in the space-time) and a given list of infected people (query curves like the one shown in red in Figure \ref{fig:space-time}).

\subsection{Encoding the Locations for Privacy Preservation}
For contact tracing purposes, the location traces of each user over a range of time (4-dimensional space-time data) needs to be assimilated in a database. This is likely to raise privacy concerns as mandated by various privacy regulation practices, e.g. the GDPR \citep{gdpr2018}. A possible approach to prevent any possible misuse of the true location data of real users is to apply a linear transformation of the data using random projections \cite{Andoni:2008}.   
For privacy preservation, as a part of the general workflow, we first apply
a distance preserving transformation function $\phi$ comprised of projections along random basis vectors. This is followed by application of a quantization function, $f_\delta$, of the projected values.

\para{Distance-preserving transformation.}
Let $\phi$ denote the transformation function which maps points from $\mathbb{R}^d$ to its corresponding images in $\mathbb{R}^p$, i.e.,
$\phi: \vec{w} \in \mathbb{R}^d \mapsto \vec{x} \in \mathbb{R}^p$.
%
The most common function for such transformation is the locality sensitive hash function (LSH) \citep{Andoni:2008}, which involves randomly choosing a set of $p$ basis vectors $\mathfrak{B}$, where $p$ is a parameter. Each point is then transformed by computing projections of the point along these $p$ basis vectors yielding the $p$ components of the transformed point in $\mathbb{R}^p$.
More concretely, the $i^{th}$ component of the transformed vector in $\mathbb{R}^p$ is given by 
\begin{equation}
x_i = \vec{w} .\vec{b}_i,
\label{eq:transform}
\end{equation}
where $\vec{w}$ is a (raw) data vector (e.g. the true user trajectories) in the space $\mathbb{R}^d$, and $\vec{b}_i \in \mathfrak{B}$ is the $i^{\mathrm{th}}$ basis vector.

A random basis ensures that computing the inverse function is non-tractable \citep{Andoni:2008}.
However, as per the Johnson-Lindenstrauss (JL) lemma \citep{johnson84extensionslipschitz}, it is known that this random projection based transformation of Equation \ref{eq:transform} is in fact distance preserving \citep{pmlr-v37-yi15}.
\citet{JiLYZT12} further improved the robustness of this distance preserving transformation by applying orthogonalization on the randomly chosen basis vectors with the help of the Gram-Schmidt method. In this paper, we specifically use the orthogonal basis vector based approach of \cite{JiLYZT12} as a definition of the transformation function $\phi$.

\para{Quantizing the projections.}
The purpose of quantization of the projected values is two fold. First, quantizing the projected values adds a further layer of obfuscation on the projected values.
Second, it helps to reduce the storage space (4 or 8 bytes of floating point vs. a single byte which allows for up to 256 possible quantized values) and hence allows faster loading of parts of the index into the main memory thereby speeding up the retrieval process.

\begin{figure}[t]
  \centering
  \includegraphics[width=.5\columnwidth]{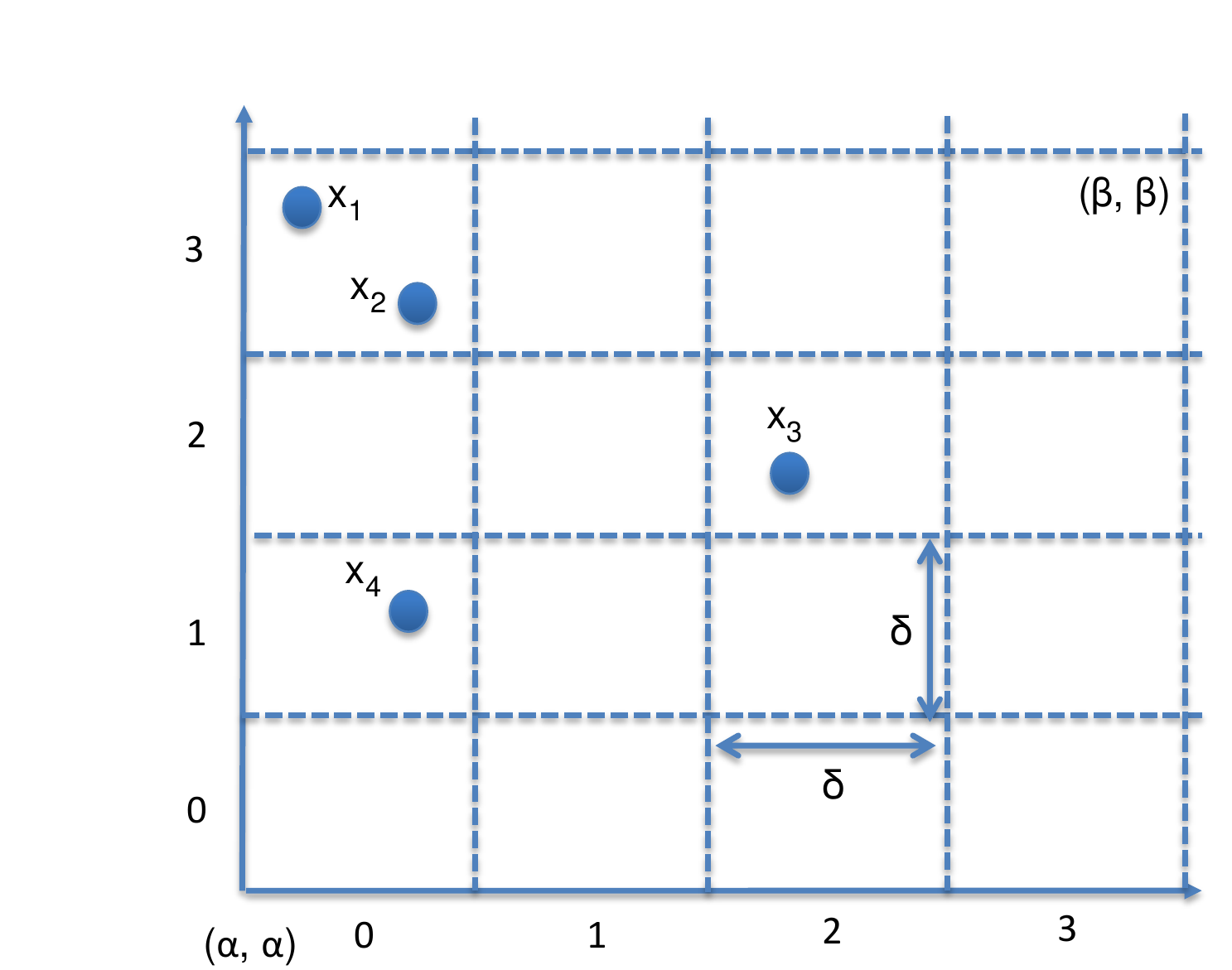}
  \caption{Decomposing $\mathbb{R}^p$ into a set of $l_{\infty}$ balls.}
  \label{fig:grid}
\end{figure}

The key idea in quantization is to transform the real-valued Cartesian space,
$\mathbb{R}^p$, into a set of non-overlapping axis-parallel grids.
More formally, each grid represents an $l_{\infty}$ ball of some positive radius $\delta \in \mathbb{R}$, taking the shape of a hyper-cube of length $\delta$.
This transformation is visualized for the particular case of $2$ dimensions in Figure \ref{fig:grid}, where each $l_{\infty}$ ball manifests itself as a square cell. Figure \ref{fig:grid} shows $4$ points, $\vec{x_1}, \ldots, \vec{x_4}$ in a two dimensional space.
%
%
If $X=\bigcup_{i=1}^{N}\{\vec{x_i}\}$ denotes
a set of $N$ points in $\mathbb{R}^p$,
to place the grid over $X$,
we first calculate the length of each grid, denoted by $\delta$. The value of $\delta$ is a function of a) the number of equi-spaced intervals $M$, in which we would want to split each axis dimension, and b) the minimum and the maximum coordinates along the axis dimensions, denoted by $\alpha$ and $\beta$ respectively.
Thus,
\begin{equation}
\delta=\frac{\beta-\alpha}{M},\quad \alpha=\min_{i=1}^{N}\min_{j=1}^{p}\vec{x_i}_j,\quad
\beta=\max_{i=1}^{N}\max_{j=1}^{p}\vec{x_i}_j
\label{eq:bounds}
\end{equation}
The $l_{\infty}$ balls are hence centred at points $\vec{c} \in \mathbb{R}^p$, where 
\begin{equation}
\vec{c}=\{\alpha + (r+\frac{1}{2})\delta\}^p, \quad r=0,\ldots,M-1
\label{eq:intervals}
\end{equation}

We then define a transformation function, $f_\delta(\vec{x})$, which represents a point $\vec{x}$ by the coordinates of its discrete grid locations along each dimension. More formally,
\begin{equation}
f_\delta(x_i) = \ceil*{\frac{x_i - \alpha}{\delta}}, \quad \forall i=1,\ldots,p.
\label{eq:transformation}
\end{equation}
The distance between two quantized points is given by
\begin{equation}
\mathcal{D}_\delta(f_\delta(\vec{x}), f_\delta(\vec{y}))=
\Big(\sum_{i=1}^{p}
\big(\ceil*{\frac{x_i - \alpha}{\delta}}-\ceil*{\frac{y_i - \alpha}{\delta}}\big)^2\Big)^{\frac{1}{2}}
\label{eq:dist_transform}
\end{equation}
%
%
%
\begin{figure}[t]
  \centering
  \includegraphics[width=.4\columnwidth]{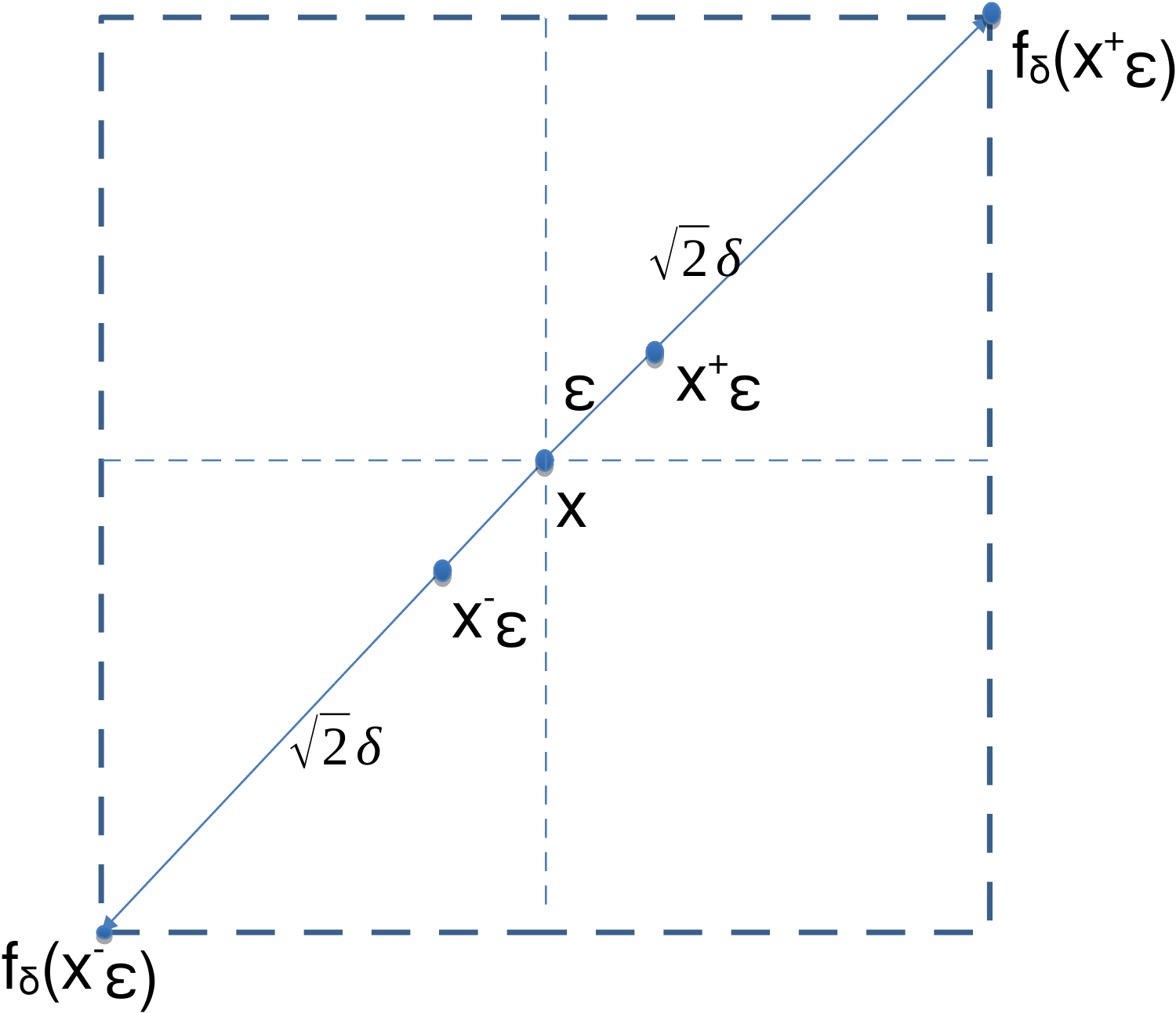}
  \caption{Maximum error in distance approximation.}
  \label{fig:distortion}
\end{figure}
%
Figure \ref{fig:distortion} demonstrates the approximation effect of the quantization in two dimensions. Maximum quantization error occurs when two points, $\vec{x_\epsilon^+}$ and $\vec{x_\epsilon^-}$, in the $\epsilon$-neighbourhood of $\vec{x}$ are transformed to two different points $f_\delta(\vec{x_\epsilon^+})$ and $f_\delta(\vec{x_\epsilon^-})$ respectively.
The separation distance between these two transformed points in two dimensions
is $\sqrt{2}\delta$, whereas for the general case of $p$ dimensions, this 
distance is $\sqrt{p}\delta$.
Hence, the maximum factor by which distances are magnified,
in the general case of $p$ dimensions, is given by
\begin{equation}
\label{eq:quantize_distorsion}
\frac{\mathcal{D}(f_\delta(\vec{x_\epsilon^+}), f_\delta(\vec{x_\epsilon^-}))}{\mathcal{D}(\vec{x_\epsilon^+}, \vec{x_\epsilon^-})}=
\frac{2\sqrt{p}\delta}{2\epsilon}=\frac{\sqrt{p}\delta}{\epsilon}
\end{equation}
As expected, this distortion can be reduced with small values of $\delta$, which
is a parameter of the quantization process. In other words, the closer a point is within the
to the corner point between two grids, i.e. lower the value of $\epsilon$,
the higher is the quantization error.

\section{Evaluation}    \label{sec:experiments}

\begin{figure}[t]
    \centering
    \includegraphics[width=0.5\columnwidth]{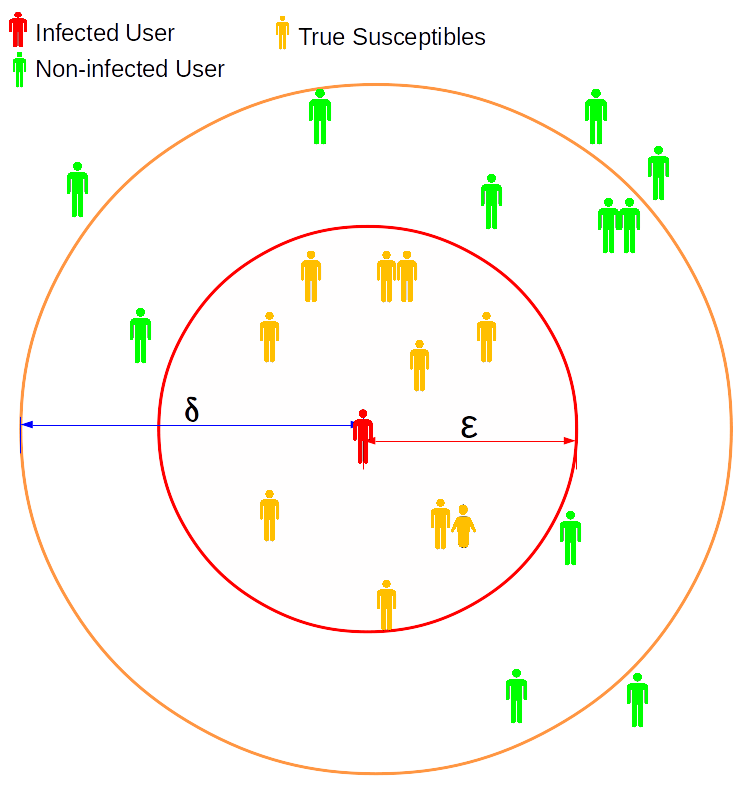}
    \caption{Simulated ghost-users (shown in amber color) corresponding to a real infected user (shown in red)}
    \label{fig:simulated_data}
\end{figure}

\begin{figure}
    \centering
    \includegraphics[width=0.49\columnwidth]{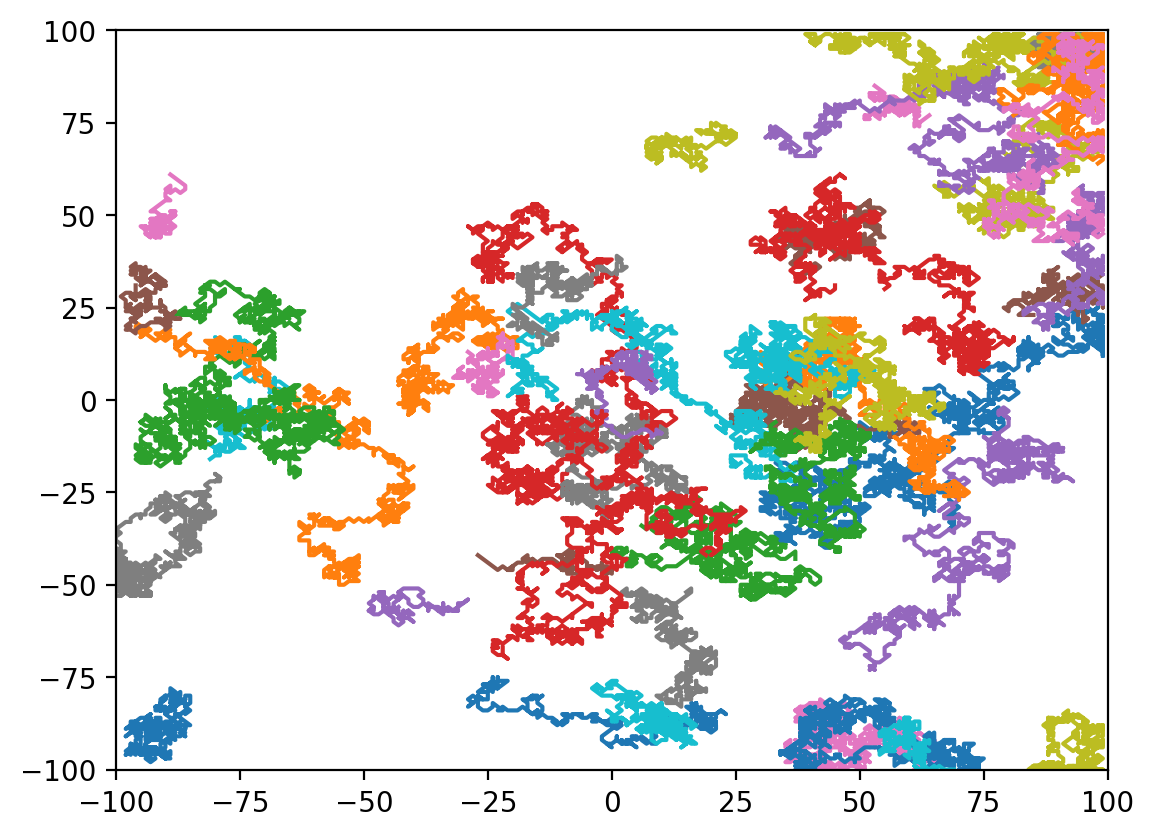}
    \includegraphics[width=0.48\columnwidth]{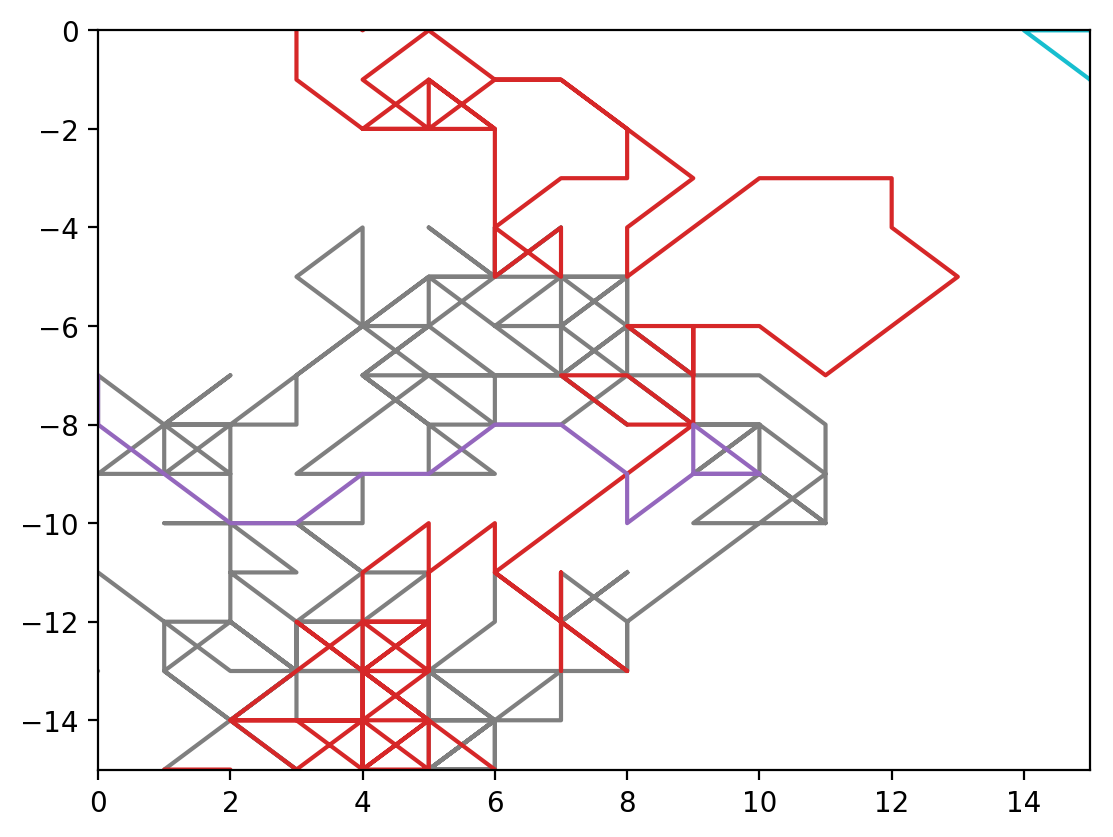}
    \caption{\emph{Left}: Random walk based trajectory data of $50$ users. \emph{Right}: A zoomed-in view for the trajectory of $3$ users. Given the red colored trajectory as a query (infected user) the objective is to retrieve the other two.}
    \label{fig:randwalk}
\end{figure}

\begin{table*}[t]
\centering
\small
\caption{ANN retrieval results for a range of different values for the number of simulated infected individuals (queries) denoted in the table as `\#Infected'. In these results, for each infected user, $r$ (\#retrieved at each time step) is set to $100$.}
\label{tab:res}
\begin{tabular}{@{}c@{~}c@{~~} c@{~~}c@{~~}c@{~~} c@{~~}c@{~~}c@{~~} c@{~~}c@{~}c@{~~}c@{~~} c@{~}c@{~}c@{~~}c@{~~}}
    \toprule
    & & \multicolumn{3}{c}{HNSW} & \multicolumn{3}{c}{KD-tree} & \multicolumn{4}{c}{PP-HNSW} & \multicolumn{4}{c}{PP-KD-tree} \\
    \cmidrule(r){3-5} \cmidrule(r){6-8} \cmidrule(r){9-12} \cmidrule(r){13-16}
    Dataset & \#Infected &
    $p$ & Time (ms) & Recall &
    $p$ & Time (ms) & Recall &
    $p$ & $M$ & Time (ms) & Recall & 
    $p$ & $M$ & Time (ms) & Recall\\
    \midrule
    \multirow{4}{*}{Traject-10K} 
    & 100 (1\%) & 4 & 1.95 & 0.9805 & 4 & 41.77 & 0.9822 & 16 & 128 & 1.18 & 0.9276 & 16 & 128 & 52.98 & 0.9824 \\
    & 200 (2\%) & 4 & 1.84 & 0.9804 & 4 & 40.58 & 0.9823 & 16 & 128 & 1.19 & 0.9284 & 16 & 128 & 53.80 & 0.9830 \\
    & 300 (3\%) & 4 & 1.83 & 0.9809 & 4 & 42.58 & 0.9822 & 16 & 128 & 1.19 & 0.9293 & 16 & 128 & 50.80 & 0.9825 \\
    & 400 (4\%) & 4 & 1.94 & 0.9797 & 4 & 42.33 & 0.9808 & 16 & 128 & 1.20 & 0.9282 & 16 & 128 & 53.47 & 0.9815 \\
    \midrule
    \multirow{4}{*}{Traject-100K} 
    & 1K (1\%) & 4 & 2.29 & 0.9587 & 4 & 42.18 & 0.9728 & 16 & 128 & 1.71 & 0.8550 & 16 & 128 & 70.34 & 0.9678 \\
    & 2K (2\%) & 4 & 2.30 & 0.9573 & 4 & 42.12 & 0.9734 & 16 & 128 & 1.72 & 0.8577 & 16 & 128 & 71.29 & 0.9686 \\
    & 3K (3\%) & 4 & 2.28 & 0.9586 & 4 & 42.85 & 0.9735 & 16 & 128 & 1.72 & 0.8580 & 16 & 128 & 71.00 & 0.9693 \\
    & 4K (4\%) & 4 & 2.20 & 0.9575 & 4 & 42.33 & 0.9729 & 16 & 128 & 1.63 & 0.8561 & 16 & 128 & 65.69 & 0.9678 \\
    \midrule
    \multirow{4}{*}{Traject-1M} 
    & 10K (1\%) & 4 & 34.04 & 0.4542 & 4 & 63.05 & 0.7952 & 16 & 1024 & 25.05 & 0.4338 & 16 & 1024 & 196.47 & 0.7895 \\
    & 20K (2\%) & 4 & 34.04 & 0.4539 & 4 & 64.36 & 0.7949 & 16 & 1024 & 26.55 & 0.4337 & 16 & 1024 & 190.45 & 0.7890 \\
    & 30K (3\%) & 4 & 32.59 & 0.4541 & 4 & 61.26 & 0.7950 & 16 & 1024 & 24.38 & 0.4341 & 16 & 1024 & 186.25 & 0.7896 \\
    & 40K (4\%) & 4 & 30.18 & 0.4541 & 4 & 91.95 & 0.7950 & 16 & 1024 & 20.09 & 0.4378 & 16 & 1024 & 218.73 & 0.7892 \\
    \midrule
    \multirow{4}{*}{CheckIn-24M} 
    & 2670 (1\%) & 4 & 0.02 & 0.9786 & 4 & 0.24 & 1.00 & 16 & 1M & 0.03 & 0.4455 & 16 & 1M & 0.61 & 0.5020 \\
    & 5340 (2\%) & 4 & 0.02 & 0.9795 & 4 & 0.24 & 1.00 & 16 & 1M & 0.03 & 0.4510 & 16 & 1M & 0.61 & 0.5016 \\
    & 8010 (3\%) & 4 & 0.02 & 0.9789 & 4 & 0.24 & 1.00 & 16 & 1M & 0.03 & 0.4520 & 16 & 1M & 0.62 & 0.5029 \\
    & 10680 (4\%) & 4 & 0.02 & 0.9796 & 4 & 0.24 & 1.00 & 16 & 1M & 0.03 & 0.4508 & 16 & 1M & 0.61 & 0.5037 \\
\bottomrule
\end{tabular}
\end{table*}

\subsection{Dataset}  \label{ss:dataset}

To study the effectiveness of our system we perform a number of experiments with both real and synthetic datasets. As a real dataset, we use the FourSquare\footnote{\url{https://drive.google.com/file/d/0BwrgZ-IdrTotZ0U0ZER2ejI3VVk/view}} global check-in dataset.
To conduct experiments on a yet larger collection, we simulate synthetic trajectories, with a different number of users (simulated agents) and number of time steps (range of time). Table \ref{tab:dataset_summary} summarizes these datasets \footnote{A prototype of the implementation and data preparation is available for research purposes at \url{https://github.com/chandanbiswas08/infectracer}.}.





\para{Simulated Ground-truths for FourSquare Check-ins}

The real FourSquare check-in data is not directly applicable for our study because the data contains only a very small number of simultaneous check-ins of two FourSquare users in the same location (a point-of-interest, e.g. a museum/restaurant).
However, to evaluate contact tracing effectiveness under laboratory-settings, our data requires to have users that came in close contact with each other (in terms of both space and time).

\begin{figure*}[t]
    \centering
    \begin{subfigure}[b]{0.24\textwidth}
        \includegraphics[width=\textwidth]{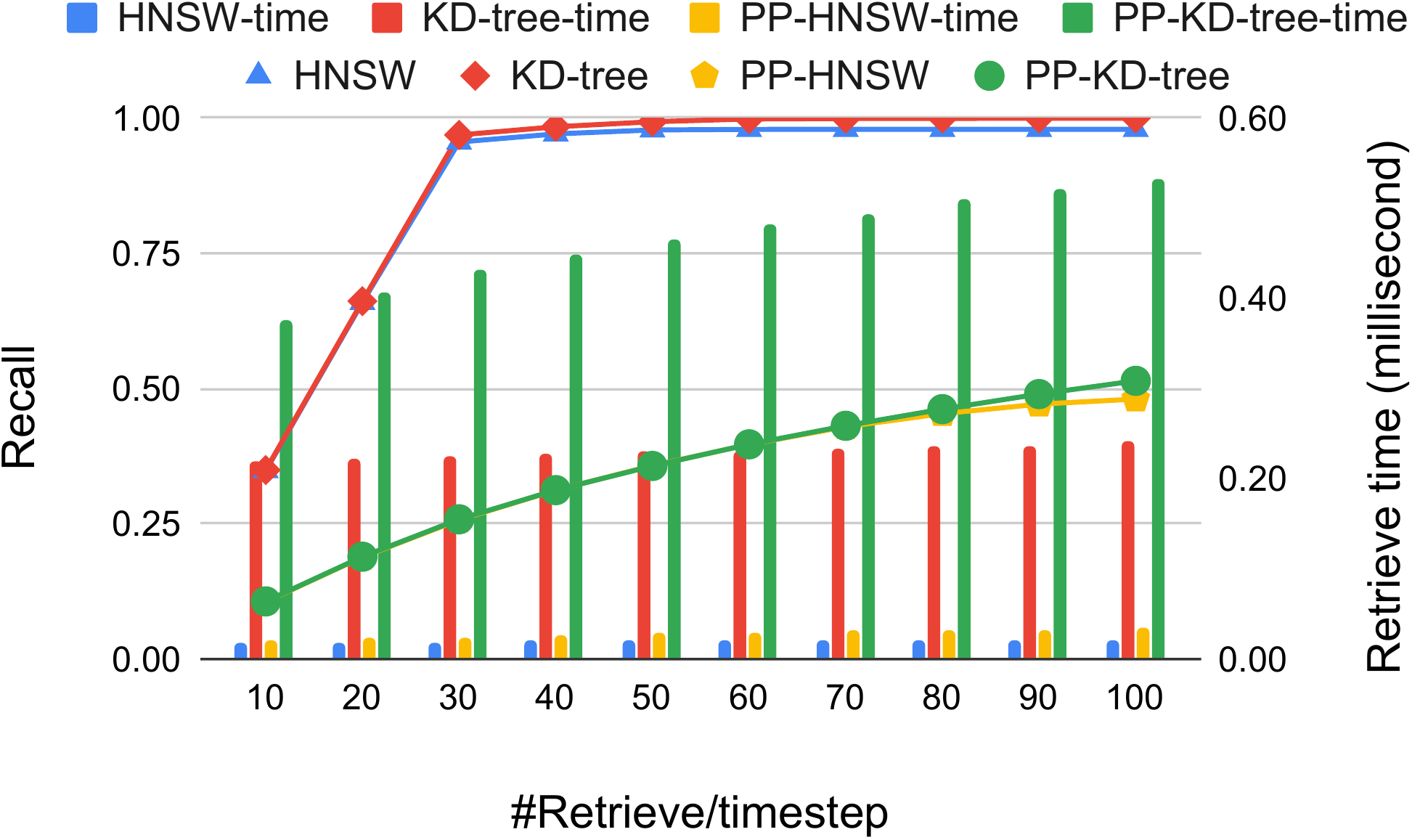}
        \caption{CheckIn-24M}
        \label{fig:CheckIn-24M_retrieve_recall}
    \end{subfigure}
    \begin{subfigure}[b]{0.24\textwidth}
        \includegraphics[width=\textwidth]{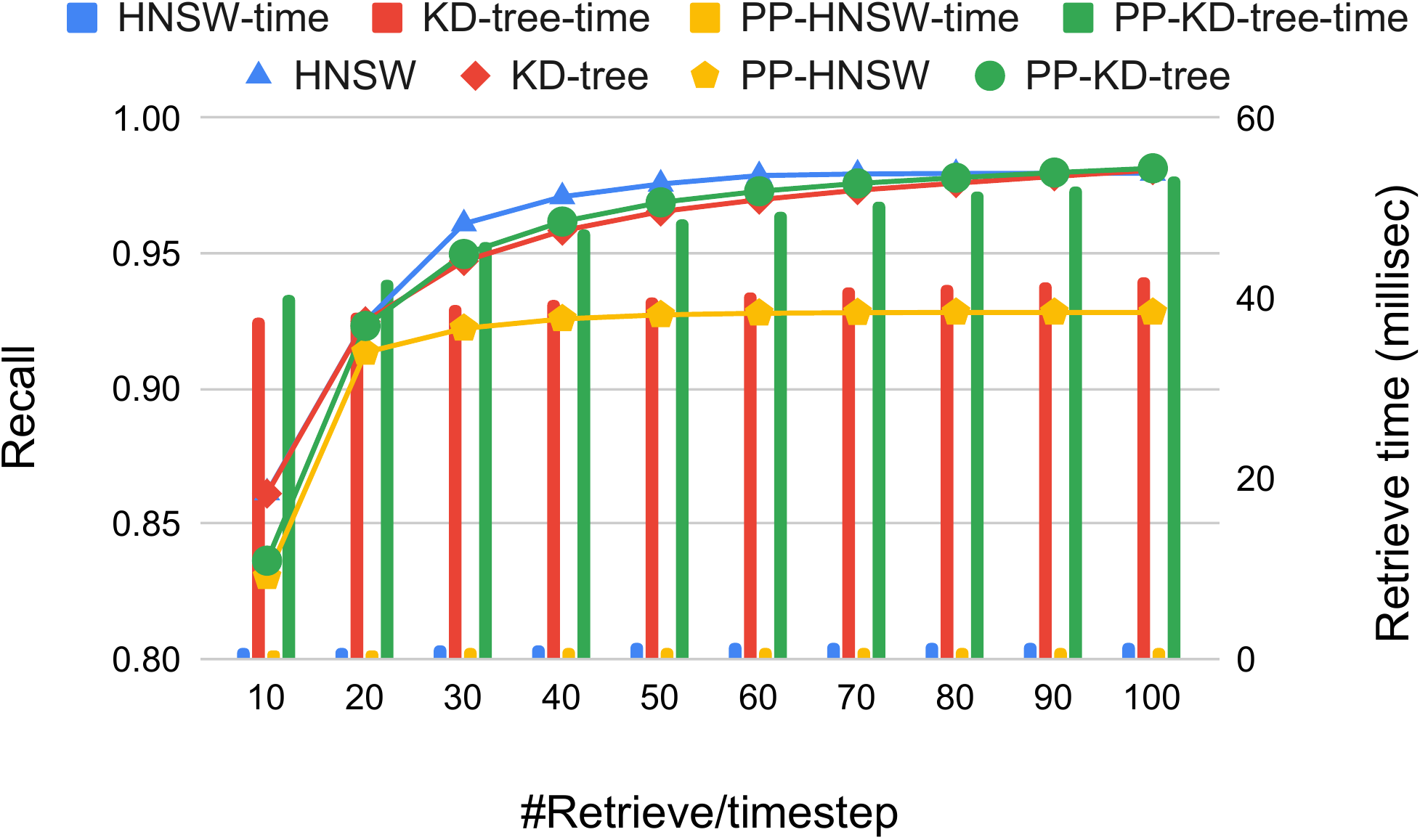}
        \caption{Traject-10K}
        \label{fig:Traject-10K_retrieve_recall}
    \end{subfigure}
    \begin{subfigure}[b]{0.24\textwidth}
        \includegraphics[width=\textwidth]{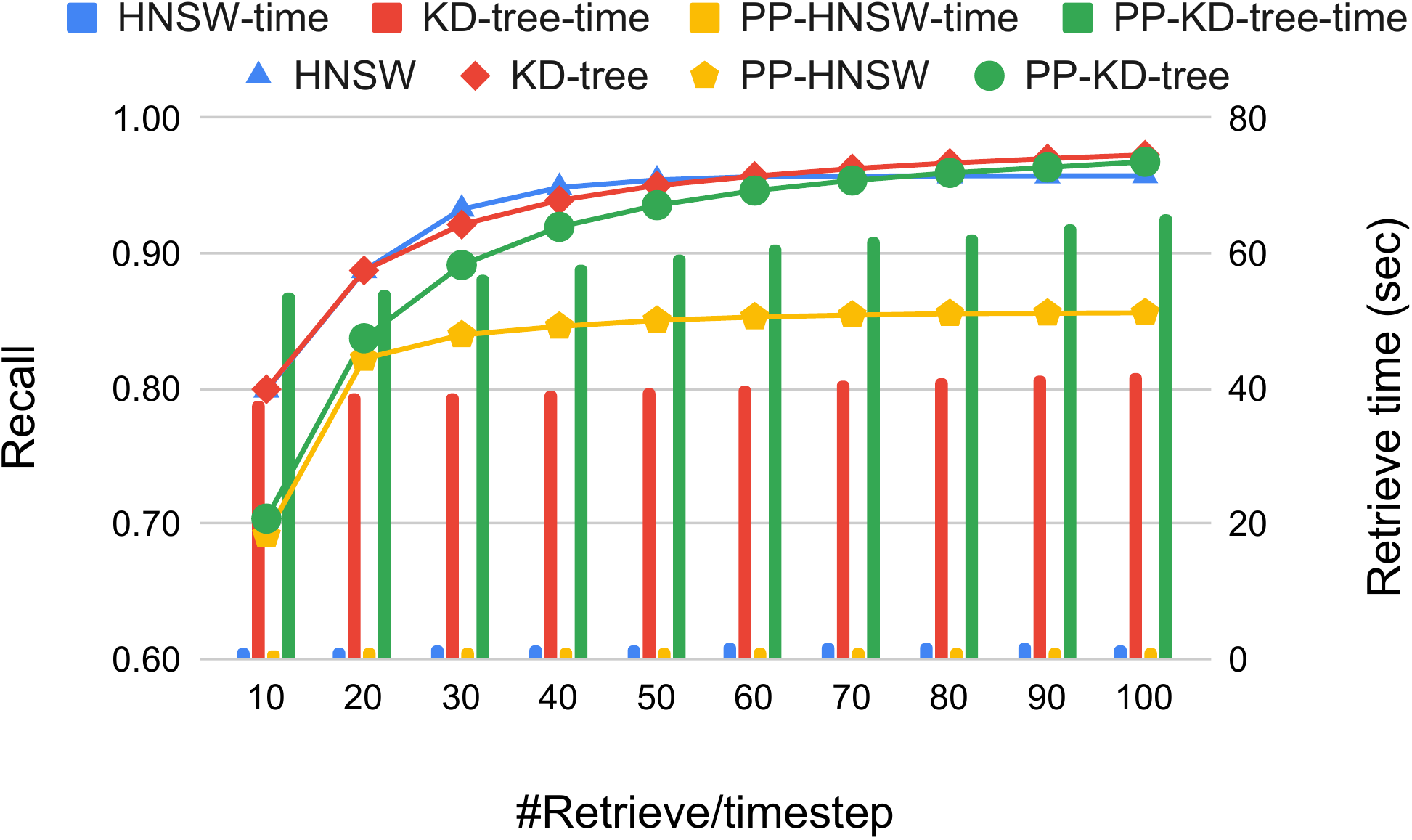}
        \caption{Traject-100K}
        \label{fig:Traject-100K_retrieve_recall}
    \end{subfigure}
    \begin{subfigure}[b]{0.24\textwidth}
        \includegraphics[width=\textwidth]{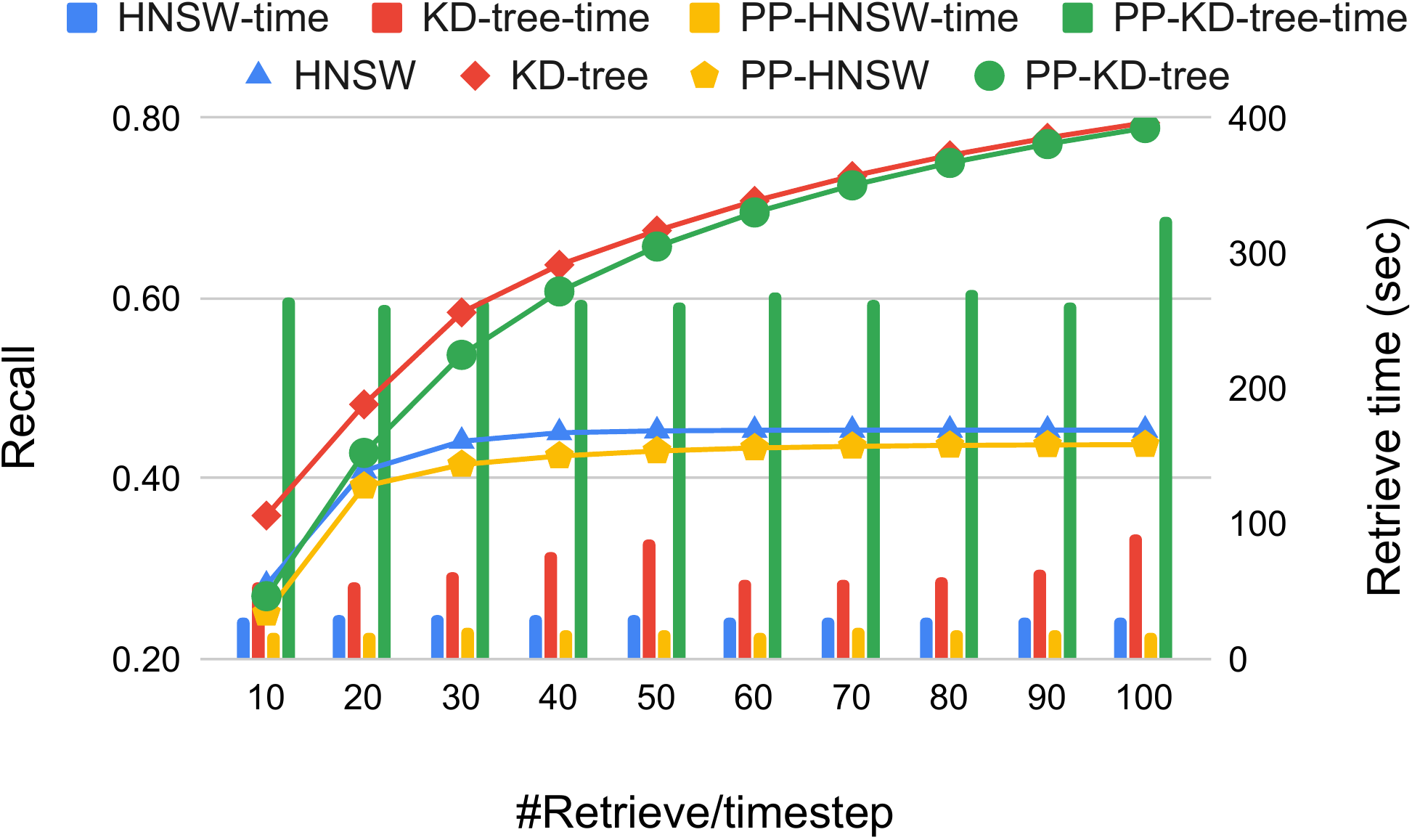}
        \caption{Traject-1M}
        \label{fig:trajectory-1M_retrieve_recall}
    \end{subfigure}
    \caption{Sensitivity of ANN retrieval effectiveness and efficiency with respect to the number of retrieved users ($r$) at each timestep. The line graph represents the recall value and bar graph represents the retrieval time for the ANN search methods.}
    \label{fig:retrieve_recall_time}
\end{figure*}

\begin{figure*}[t]
    \centering
    \begin{subfigure}[b]{0.24\textwidth}
        \includegraphics[width=\textwidth]{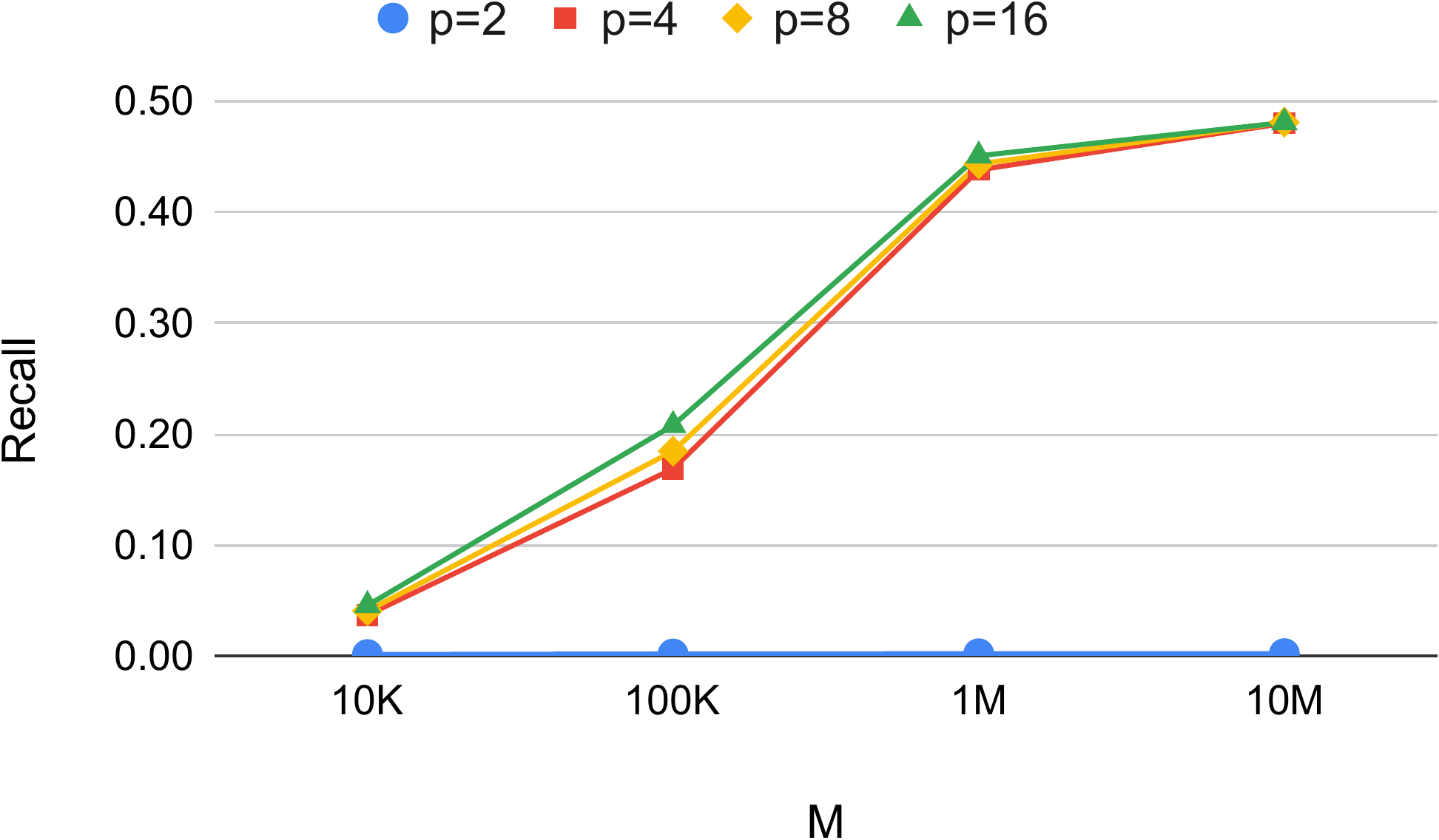}
        \caption{PP-HNSW:CheckIn-24M}
        \label{fig:CheckIn-24M_HNSW_bin_recall}
    \end{subfigure}
    \begin{subfigure}[b]{0.24\textwidth}
        \includegraphics[width=\textwidth]{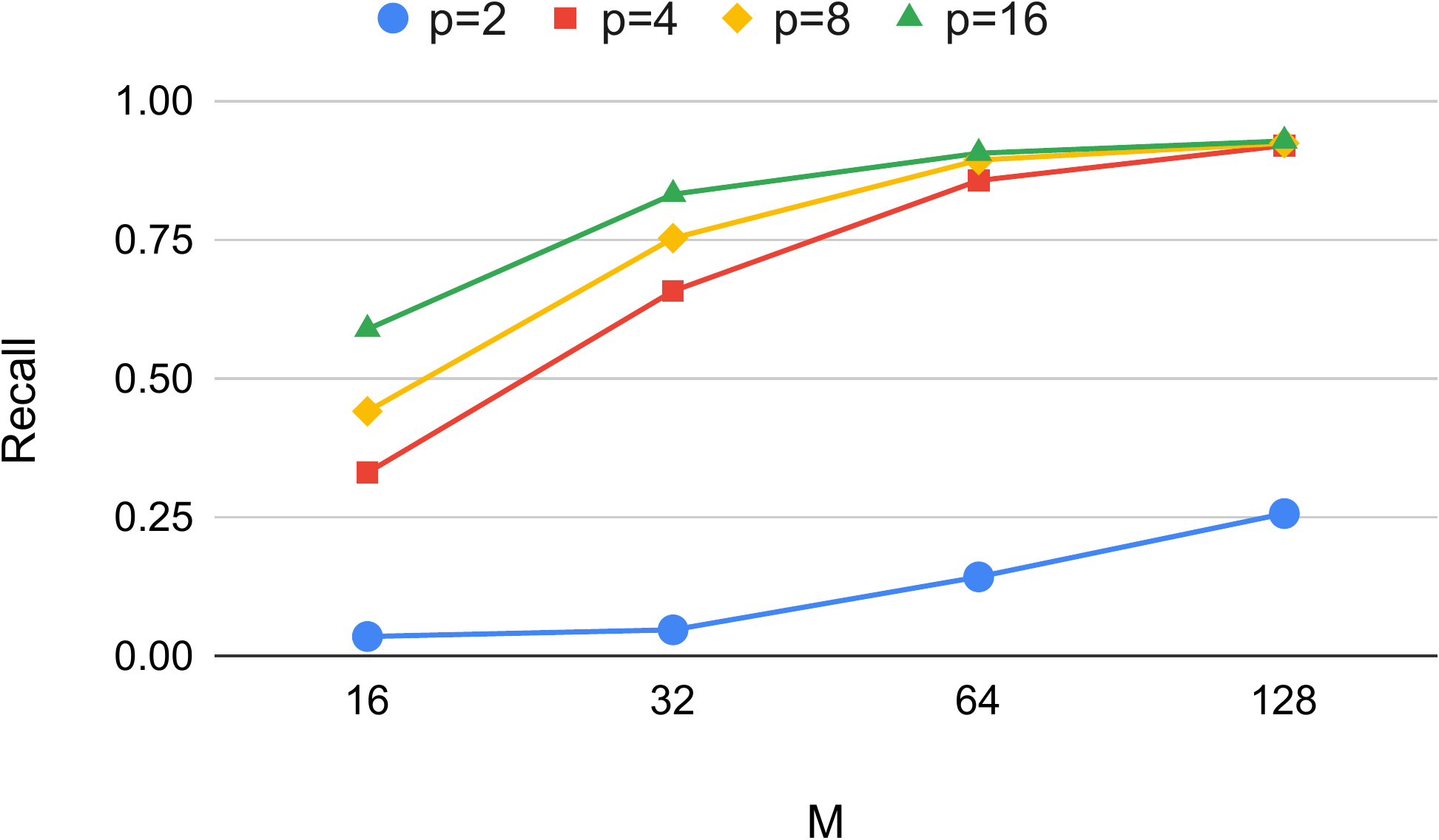}
        \caption{PP-HNSW:Traject-10K}
        \label{fig:Traject-10K_HNSW_bin_recall}
    \end{subfigure}
    \begin{subfigure}[b]{0.24\textwidth}
        \includegraphics[width=\textwidth]{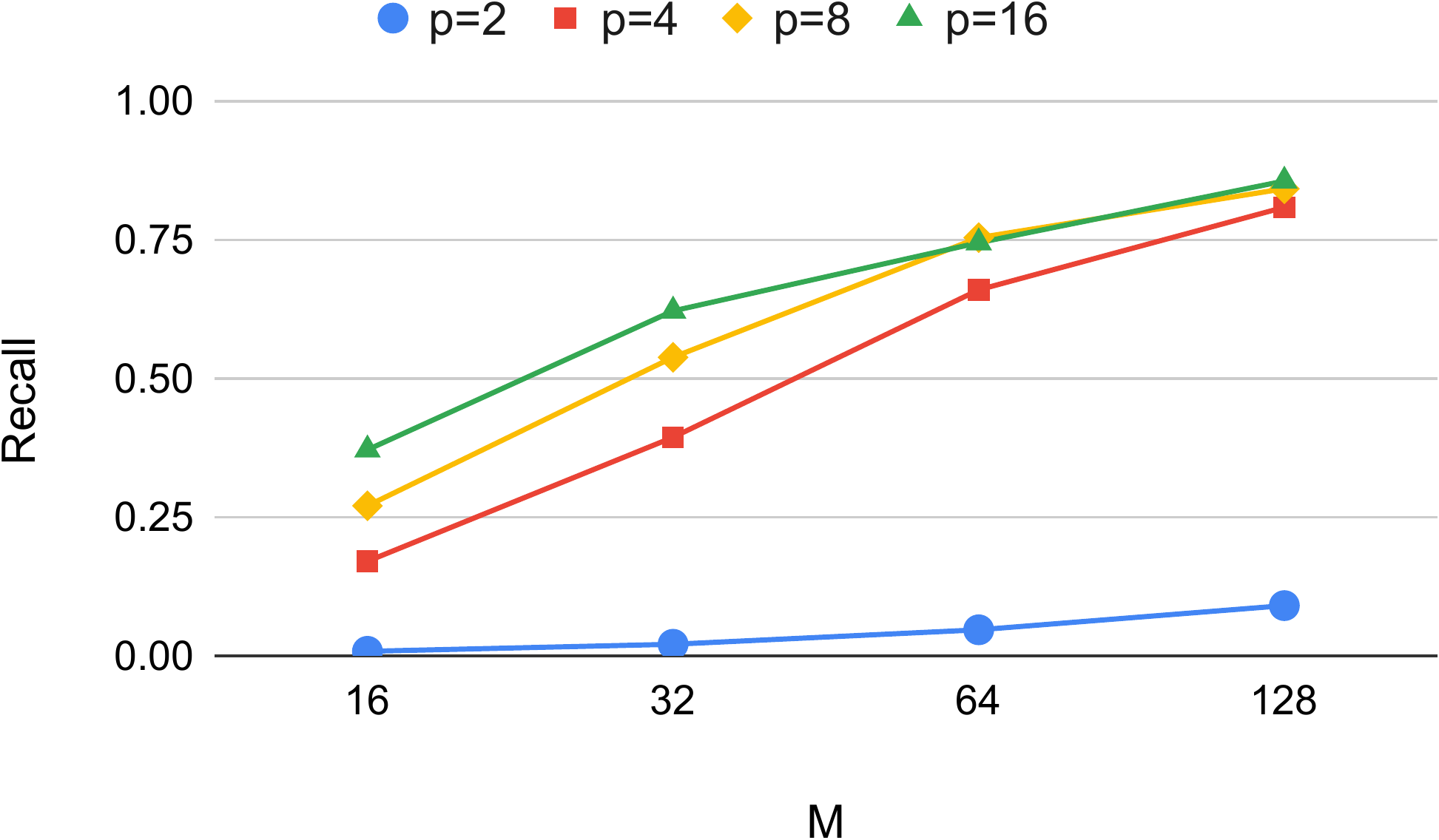}
        \caption{PP-HNSW:Traject-100K}
        \label{fig:Traject-100K_HNSW_bin_recall}
    \end{subfigure}
    \begin{subfigure}[b]{0.24\textwidth}
        \includegraphics[width=\textwidth]{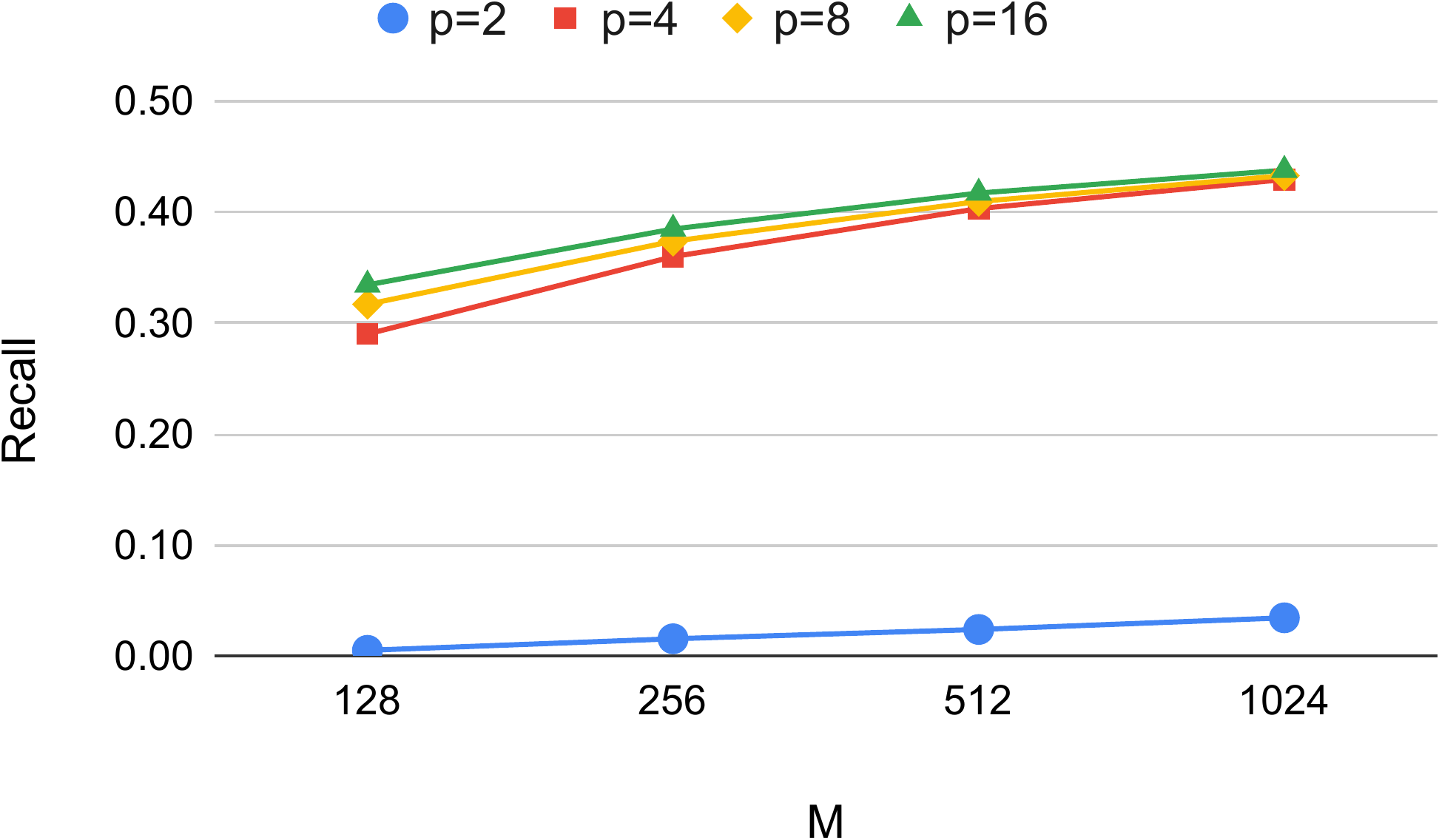}
        \caption{PP-HNSW:Traject-1M}
        \label{fig:trajectory-1M_HNSW_bin_recall}
    \end{subfigure}

    \begin{subfigure}[b]{0.24\textwidth}
        \includegraphics[width=\textwidth]{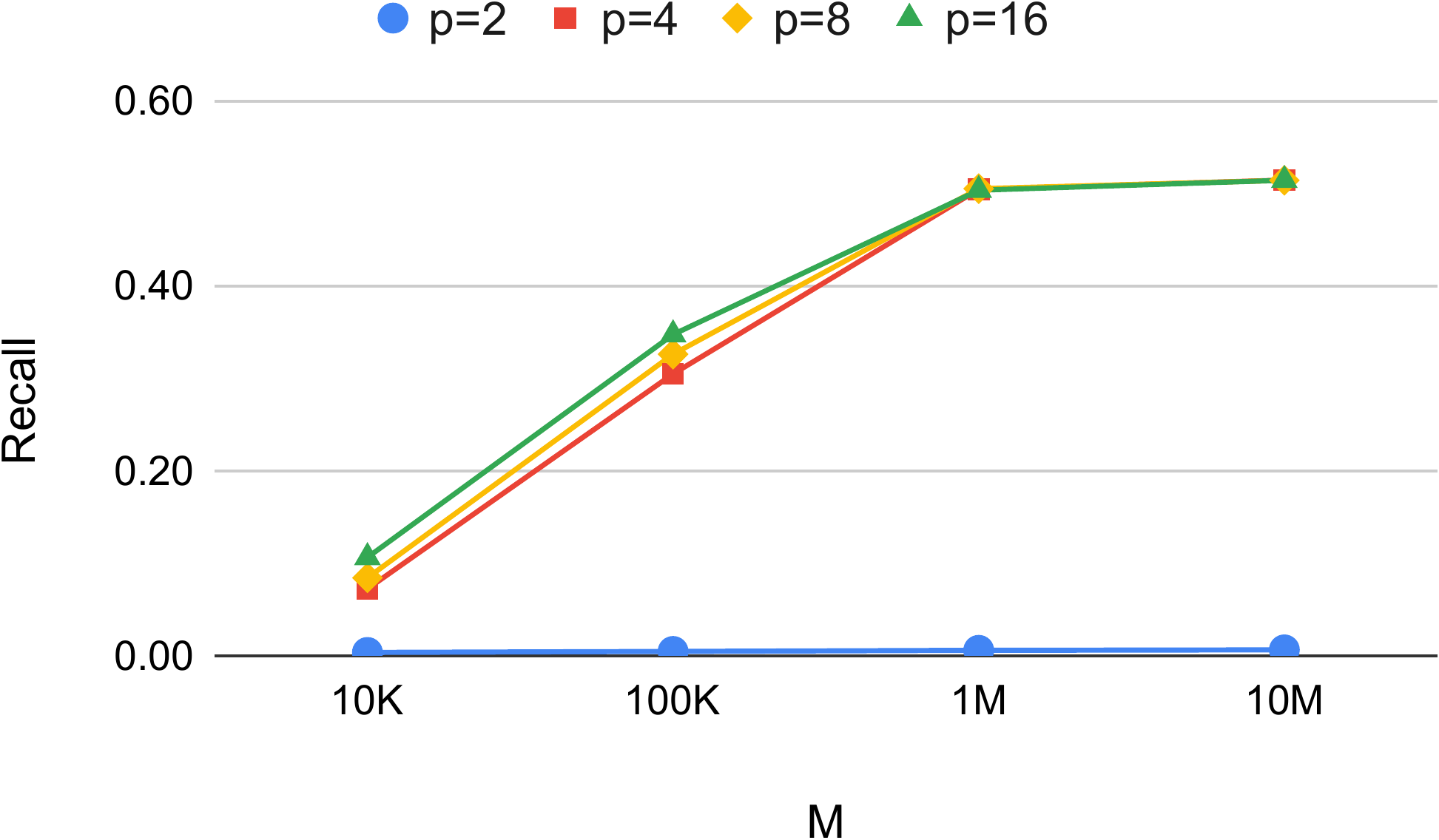}
        \caption{PP-KD-tree:CheckIn-24M}
        \label{fig:CheckIn-24M_KD-tree_bin_recall}
    \end{subfigure}
    \begin{subfigure}[b]{0.24\textwidth}
        \includegraphics[width=\textwidth]{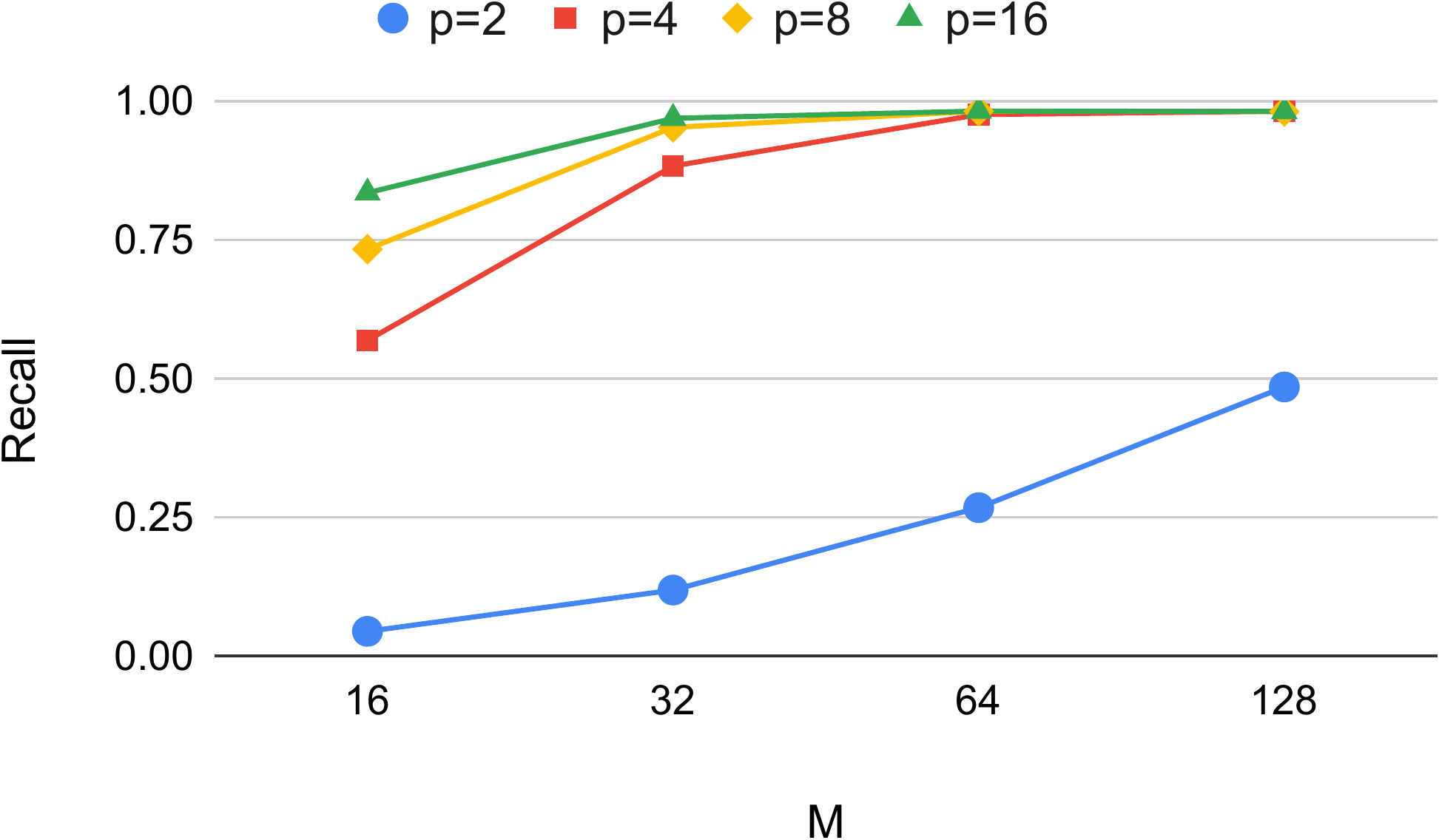}
        \caption{PP-KD-tree:Traject-10K}
        \label{fig:Traject-10K_KD-tree_bin_recall}
    \end{subfigure}
    \begin{subfigure}[b]{0.24\textwidth}
        \includegraphics[width=\textwidth]{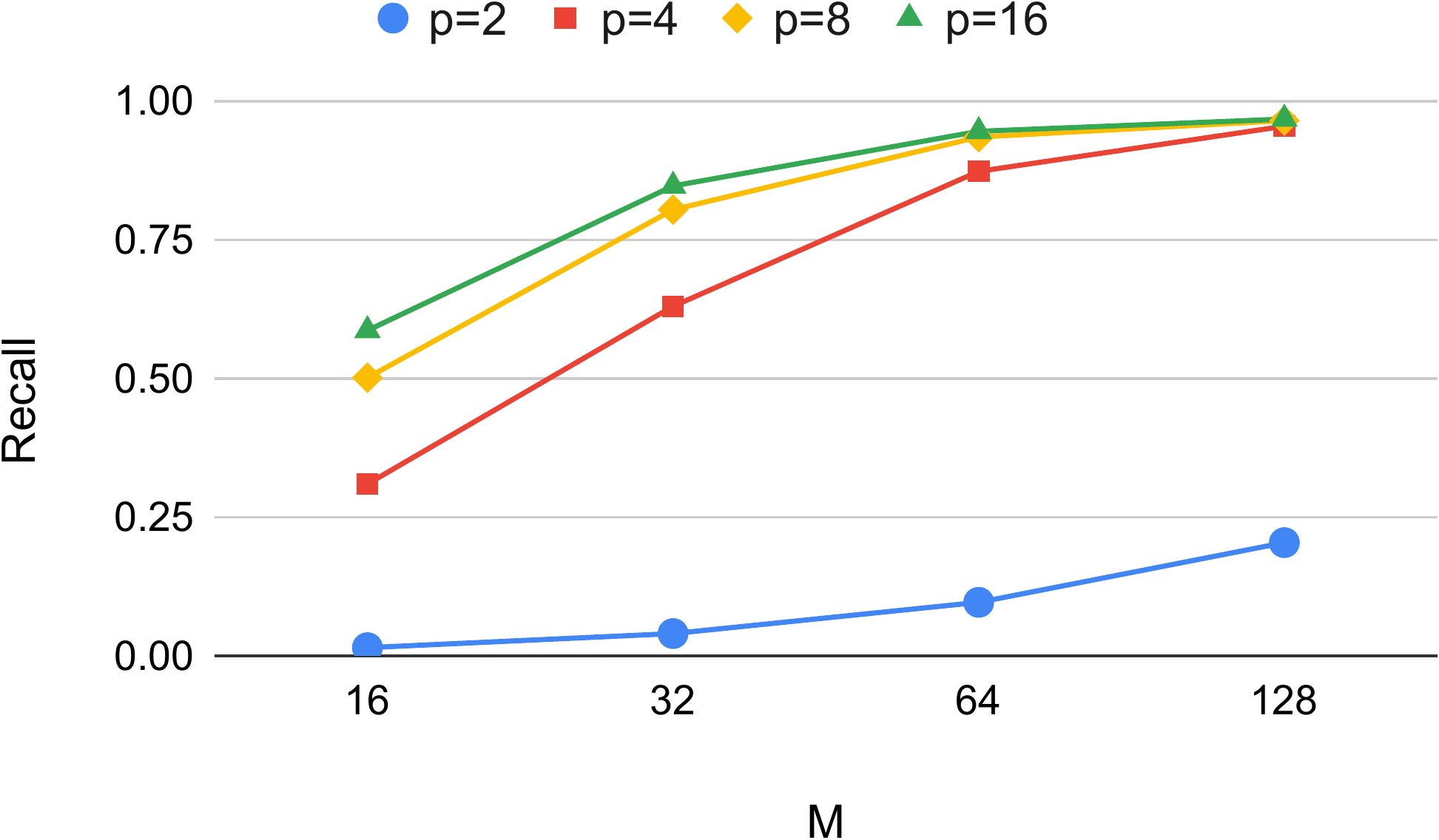}
        \caption{PP-KD-tree:Traject-100K}
        \label{fig:Traject-100K_KD-tree_bin_recall}
    \end{subfigure}
    \begin{subfigure}[b]{0.24\textwidth}
        \includegraphics[width=\textwidth]{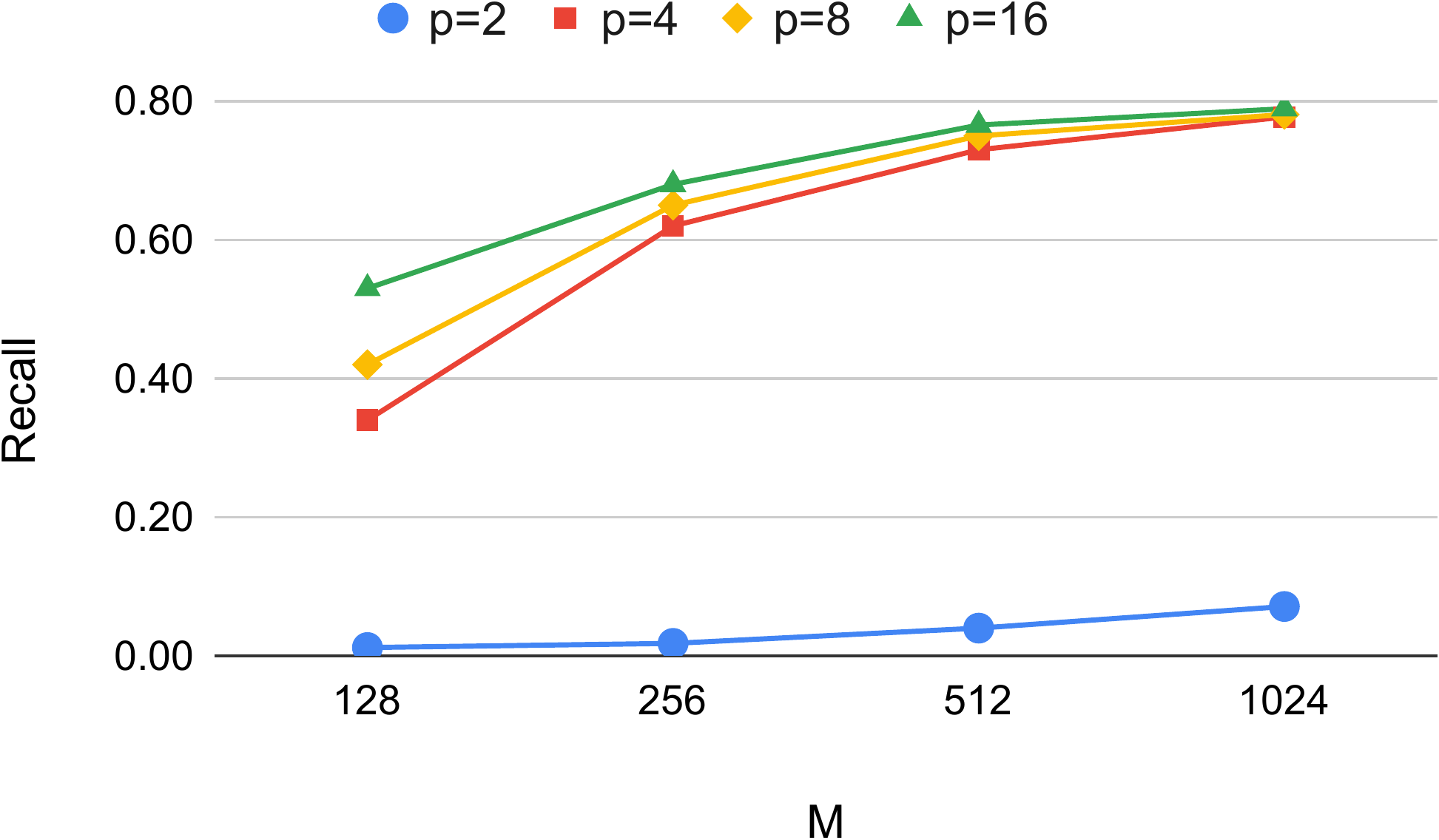}
        \caption{PP-KD-tree:Traject-1M}
        \label{fig:trajectory-1M_KD-tree_bin_recall}
    \end{subfigure}
    \caption{Sensitivity of ANN retrieval effectiveness with variations in the projection dimension ($p$) and the number of quantization intervals ($M$).}
    \label{fig:bin_recall}
\end{figure*}

As a solution, we undertake a simple simulation model to generate pseudo-user interactions (likely contacts). First, we filter the original dataset to retain only one check-in per user. This makes the simulation algorithm easier to manage.
Next, for each user $U$ (having a unique id), we generate a mutually exclusive set of `pseudo-users' or `ghost-users'. For a user $U$, as per the generation mechanism, this set of pseudo-users hence represent the ground-truth or the target set of users that need to be retrieved given the current user $U$ as a query. Note that since all the original/real user check-ins were sufficiently apart in space-time coordinates, it is likely that the neighbourhood of a user comprised of the ghost-user check-ins are also far apart (in which case one can rely with sufficient confidence on the simulated ground-truth data).
More concretely, for each user $U$ we generate $p+n$ number of ghost-users in a $\delta$ neighbourhood, out of which $p$ belong to an $\epsilon$ neighbourhood ($\epsilon<\delta$). If $U$ is infected person then the target is to retrieve the set of $p$ ghost-users.
Figure \ref{fig:simulated_data} presents a visualization of the simulated pseudo-users corresponding to a real infected user (red person in the figure). 
As particular values of $\epsilon$ and $\delta$, we use $1$ and $2$. The values of $p$ and $n$ were set to $30$ and $60$ respectively. The value of $n$ is set to be higher than that of $p$ in order to make the ANN retrieval task more challenging.

\begin{table}[t]
\centering
\small
\caption{Summary of the dataset used in our experiments. $\rho$ denotes the population density (\#users in a grid cell).}
\begin{tabular}{@{}l@{~}r@{~~}c@{~}c@{~~}c@{~}r@{}}
\toprule
Dataset & \#User & \#Instances & \#Step ($\tau$) & Grid ($\beta\times \beta$) & $\rho$ \\
\midrule
Traject-10K & 10K & 1500K & 100-200 & 100$\times$100 & 149.94\\
Traject-100K & 100K & 15M & 100-200 & 1K$\times$1K & 14.99 \\
Traject-1M & 1M & 150M & 100-200 & 1K$\times$1K & 149.97 \\
CheckIn-24M & 24M & 24M & 1 & 107$\times$337 & 670.85\\
\bottomrule
\end{tabular}
\label{tab:dataset_summary}
\end{table}



\subsubsection{Generating Synthetic Trajectory Dataset}
Since the real dataset is limited by the number of available check-ins, in order to collate a larger dataset of locations we generate synthetic data with random walk. Although the real trajectory paths of people are far from being random, the generated data despite being random serves its purpose in the context of our experiments, which is to evaluate the effectiveness of ANN on large volumes of location data.
%
%
To generate synthetic data, $N$ simulated agents are initialized each at a randomly chosen location within a 3 dimensional bounding box (each side of the bounding box being in the range $[0, \beta]$) with uniform probability.
If the location of the $i$-th agent at time-step $t$ is denoted by $(x^i_t, y^i_t, z^i_t)$, 
its location $x$ coordinate's value at the next time step is given by
\begin{equation}
\label{eq:randwalk}
x^i_{t+1} = x^i_{t} + \mathcal{U}(-1,1),\,\, x^i_{t} \in [0, \beta]
\end{equation}
and so on for the other spatial dimensions ($\beta$ is the length of each edge of the bounding cube). The process in Equation \ref{eq:randwalk} is repeated for an agent for $\tau_i$ number of steps where $\tau_{min} < \tau_i < \tau_{max}$. For our experiments, we used $\tau_{min}=100$ and $\tau_{max}=200$.
Each generated spatial location for the $i$-th agent ($\tau_i$ number of them in total) is then appended with the time dimension, yielding the set of points of the form
\begin{equation}
L_i = \cup_{t=0}^{\tau_i}\{(x^i_{t},y^i_{t},z^i_{t},t)\}.
\end{equation}

While generating the dataset at each step, if two agents are found to come sufficiently close to each other, i.e. within an $\epsilon$-neighborhood of each other ($\epsilon$ set to 1 similar to the Foursquare dataset settings), we insert each point into the ground-truth (susceptible) list of the other.
We generate different synthetic datasets with three different values of $N$ (number of simulated agents), namely $10$K, $100$K and $1$M. To name the datasets with a common prefix `Traject-' followed by the value of $N$. Table \ref{tab:dataset_summary} summarizes the datasets.
Figure \ref{fig:randwalk} shows a sample of the generated data with $50$ users for the purpose of illustration.


\begin{figure*}[t]
    \centering
    \begin{subfigure}[b]{0.24\textwidth}
        \includegraphics[width=\textwidth]{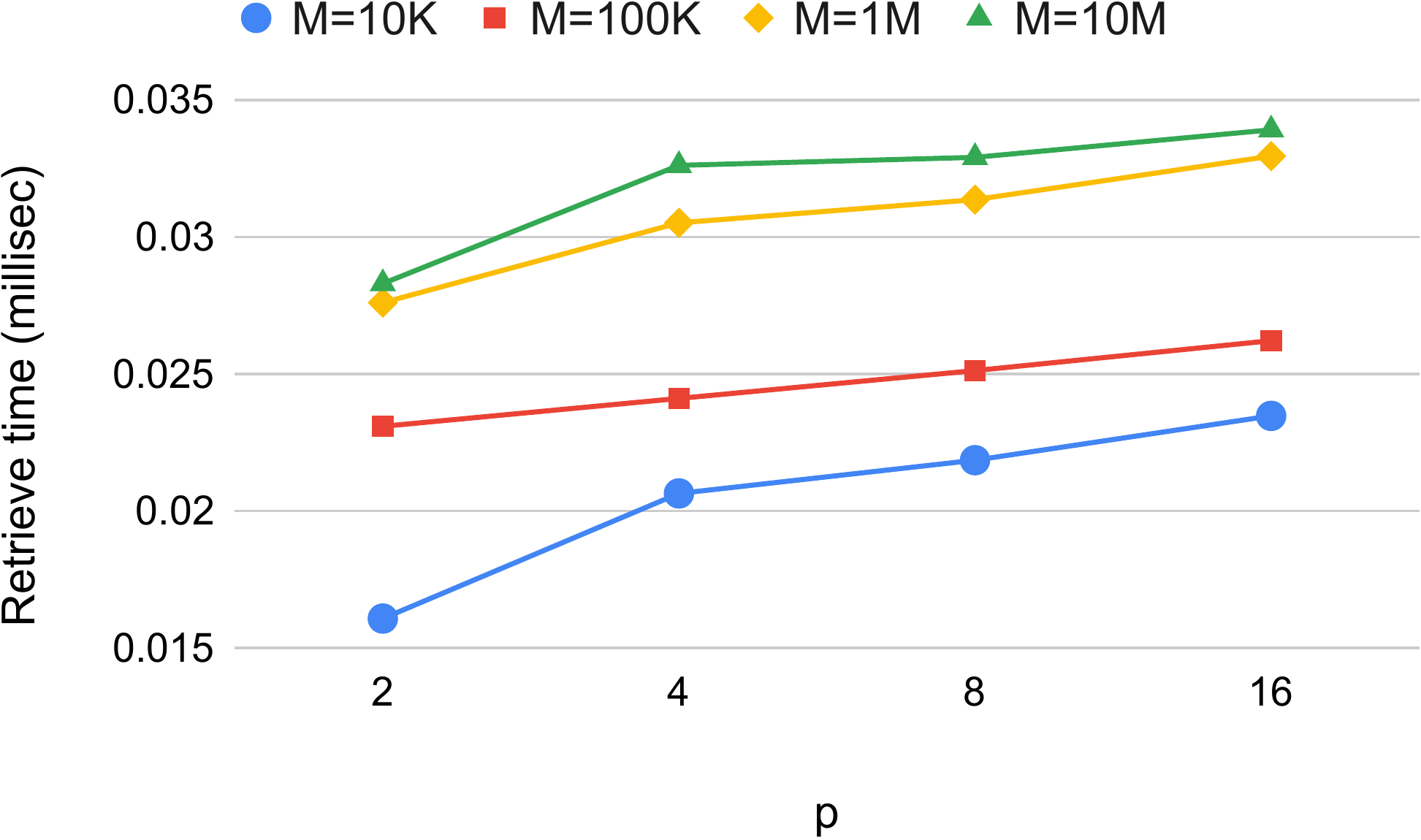}
        \caption{PP-HNSW:CheckIn-24M}
        \label{fig:CheckIn-24M_HNSW_dim_recall}
    \end{subfigure}
    \begin{subfigure}[b]{0.24\textwidth}
        \includegraphics[width=\textwidth]{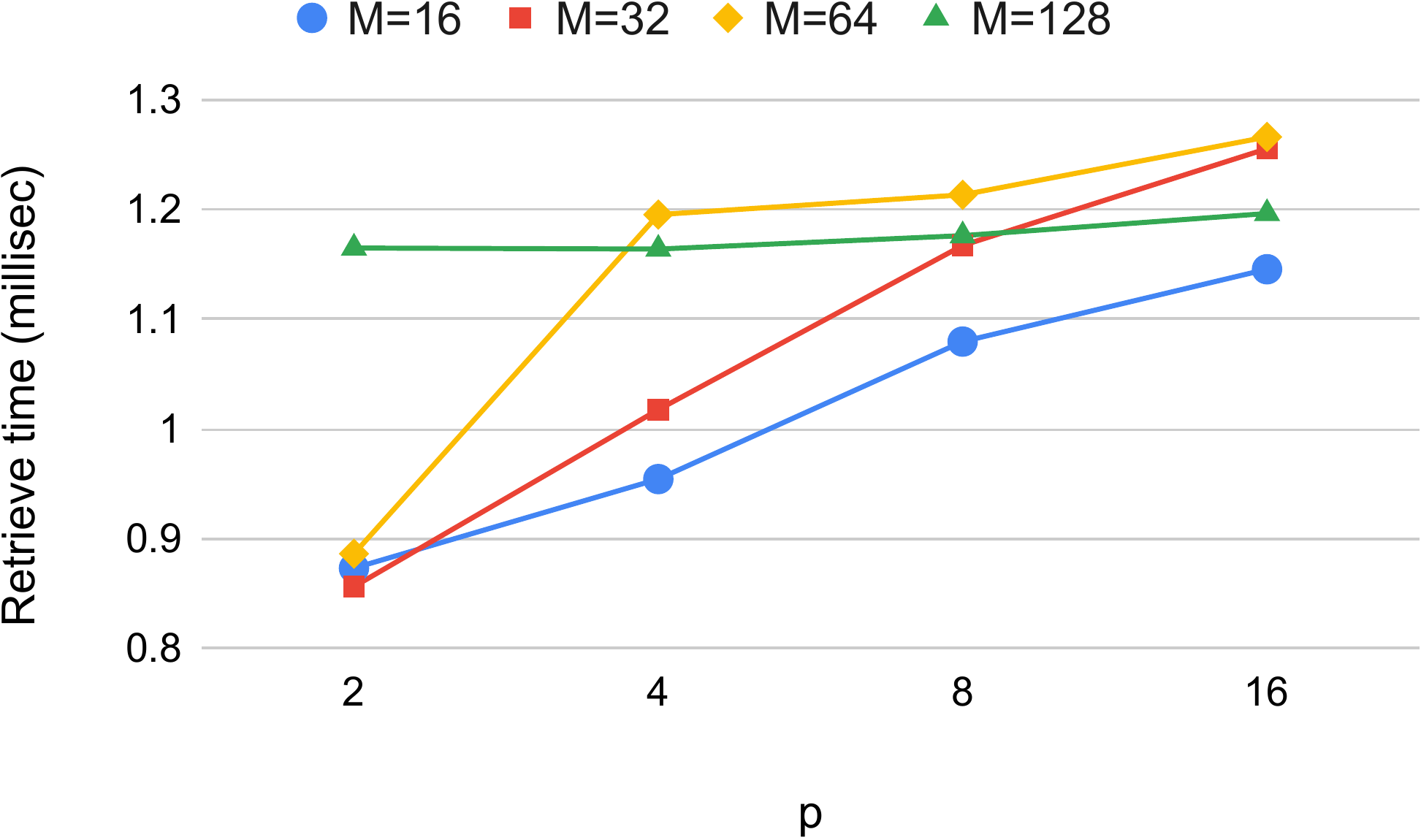}
        \caption{PP-HNSW:Traject-10K}
        \label{fig:Traject-10K_HNSW_dim_recall}
    \end{subfigure}
    \begin{subfigure}[b]{0.24\textwidth}
        \includegraphics[width=\textwidth]{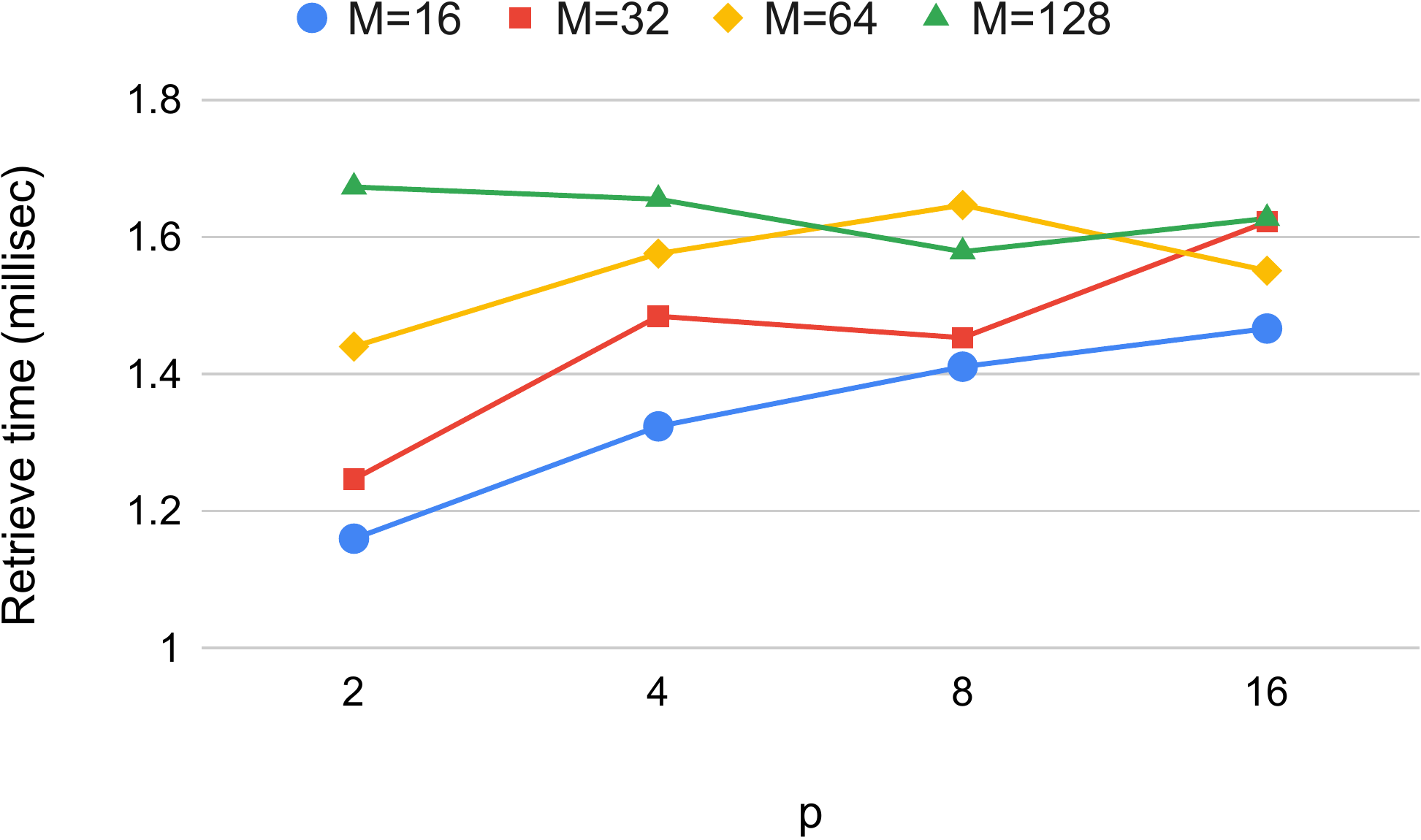}
        \caption{PP-HNSW:Traject-100K}
        \label{fig:Traject-100K_HNSW_dim_recall}
    \end{subfigure}
    \begin{subfigure}[b]{0.24\textwidth}
        \includegraphics[width=\textwidth]{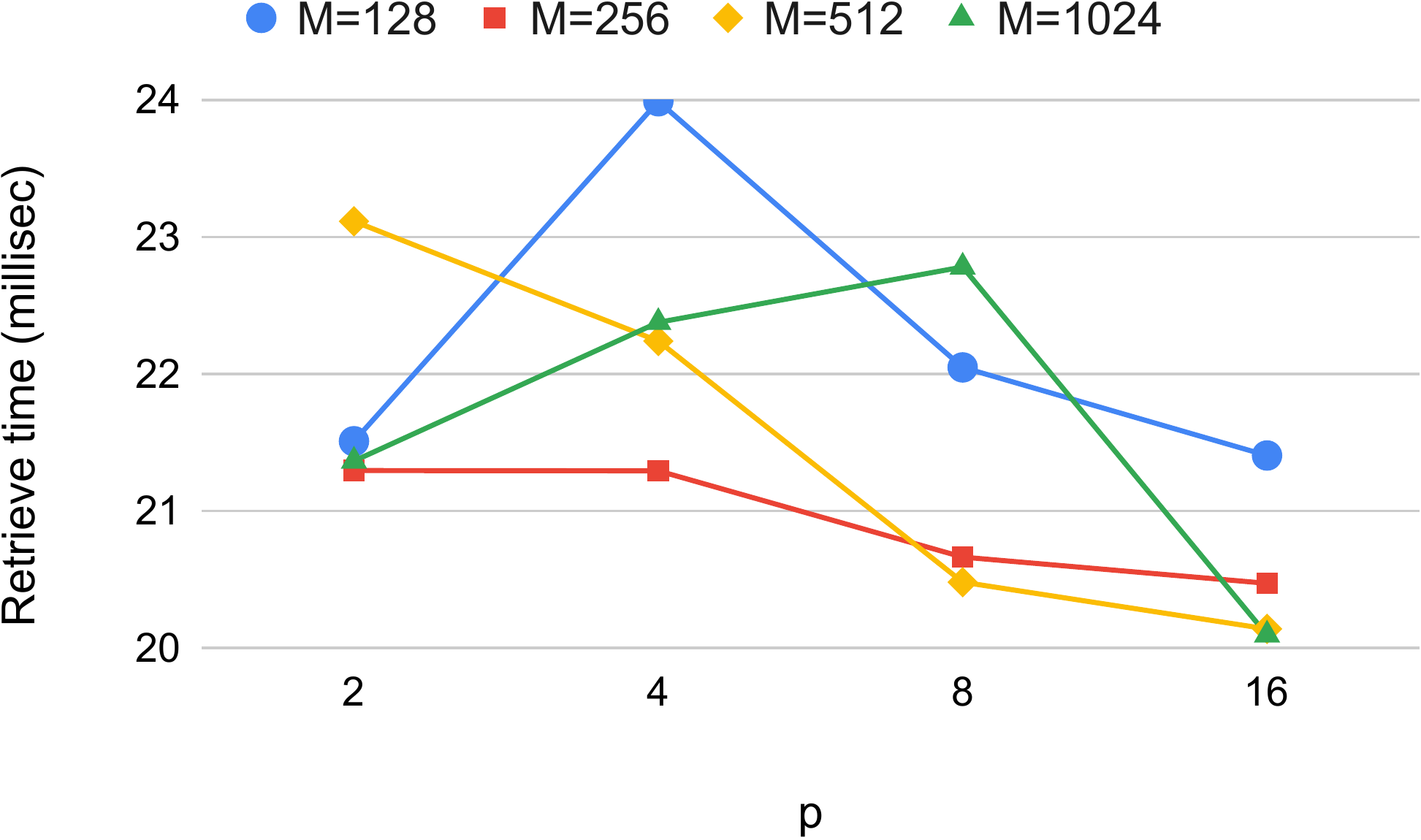}
        \caption{PP-HNSW:Traject-1M}
        \label{fig:trajectory-1M_HNSW_dim_recall}
    \end{subfigure}

    \begin{subfigure}[b]{0.24\textwidth}
        \includegraphics[width=\textwidth]{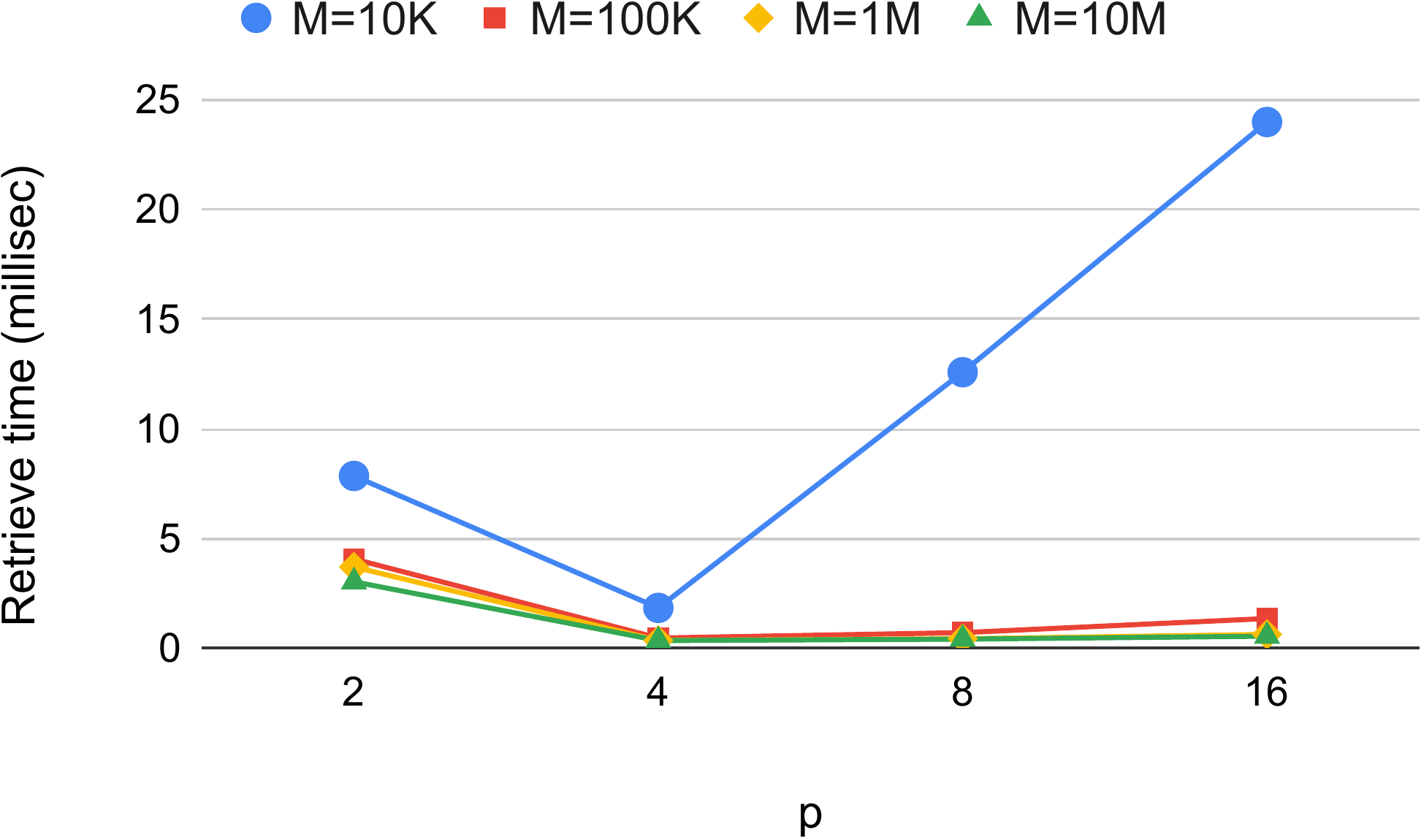}
        \caption{PP-KD-tree:CheckIn-24M}
        \label{fig:CheckIn-24M_KD-tree_dim_recall}
    \end{subfigure}
    \begin{subfigure}[b]{0.24\textwidth}
        \includegraphics[width=\textwidth]{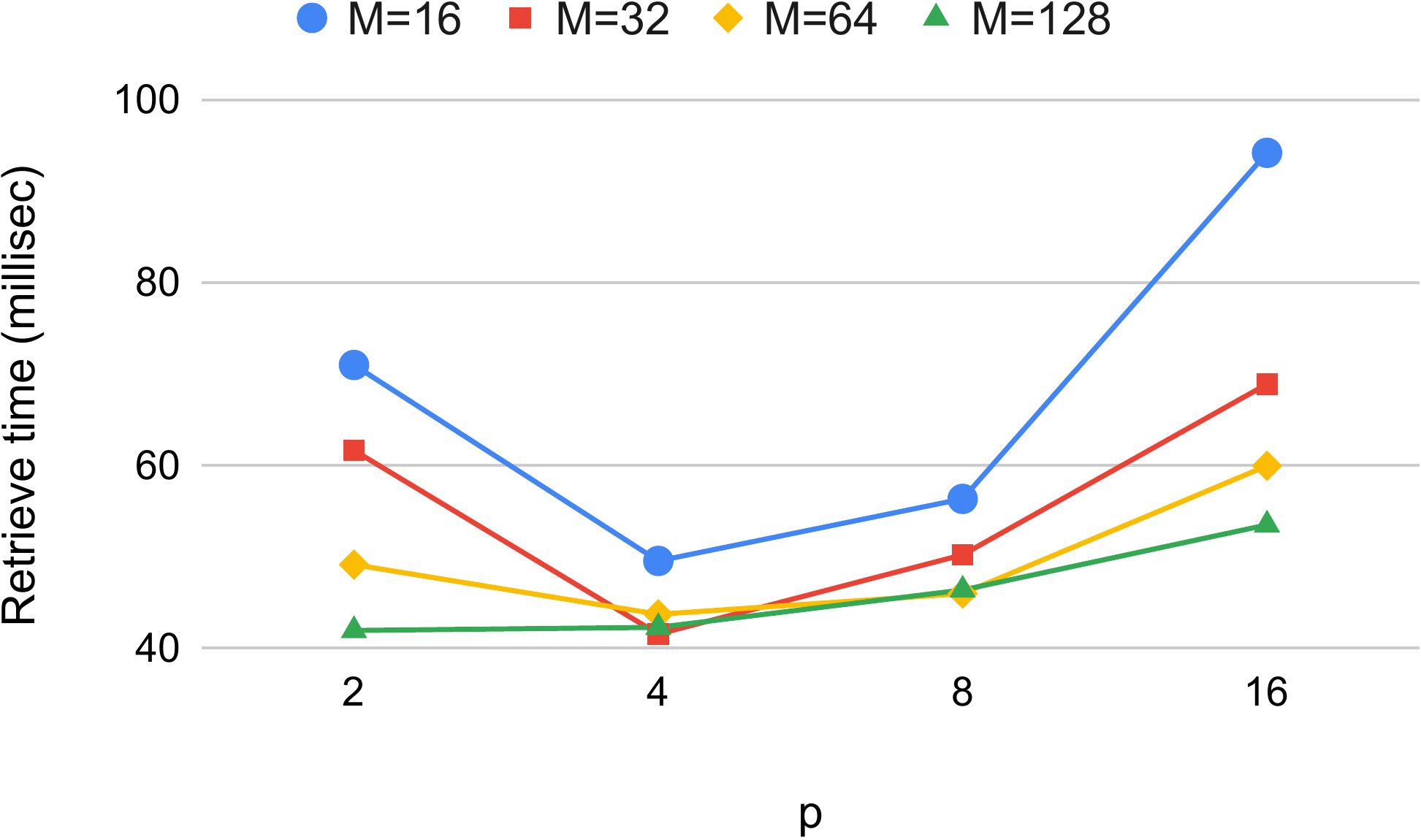}
        \caption{PP-KD-tree:Traject-10K}
        \label{fig:Traject-10K_KD-tree_dim_recall}
    \end{subfigure}
    \begin{subfigure}[b]{0.24\textwidth}
        \includegraphics[width=\textwidth]{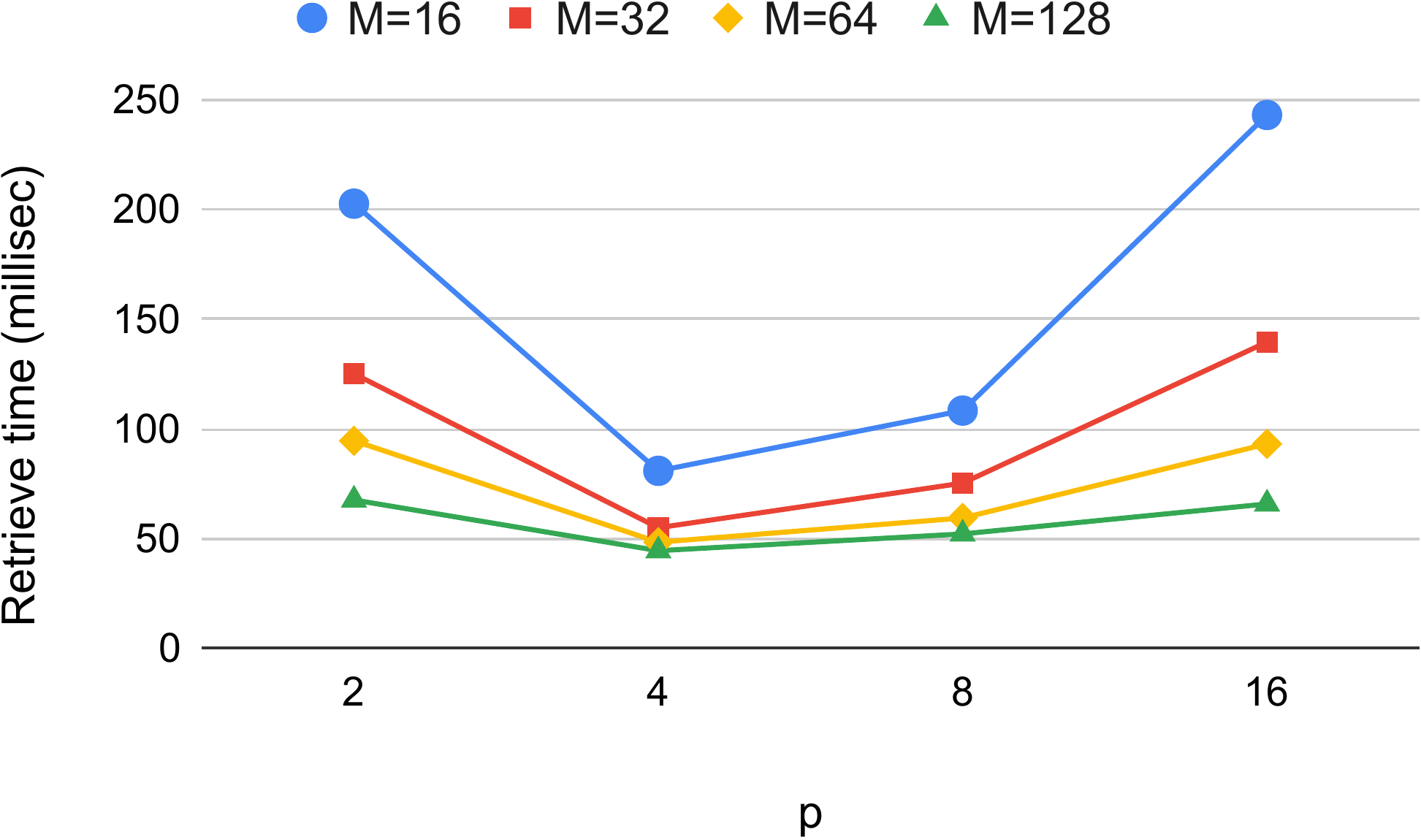}
        \caption{PP-KD-tree:Traject-100K}
        \label{fig:Traject-100K_KD-tree_dim_recall}
    \end{subfigure}
    \begin{subfigure}[b]{0.24\textwidth}
        \includegraphics[width=\textwidth]{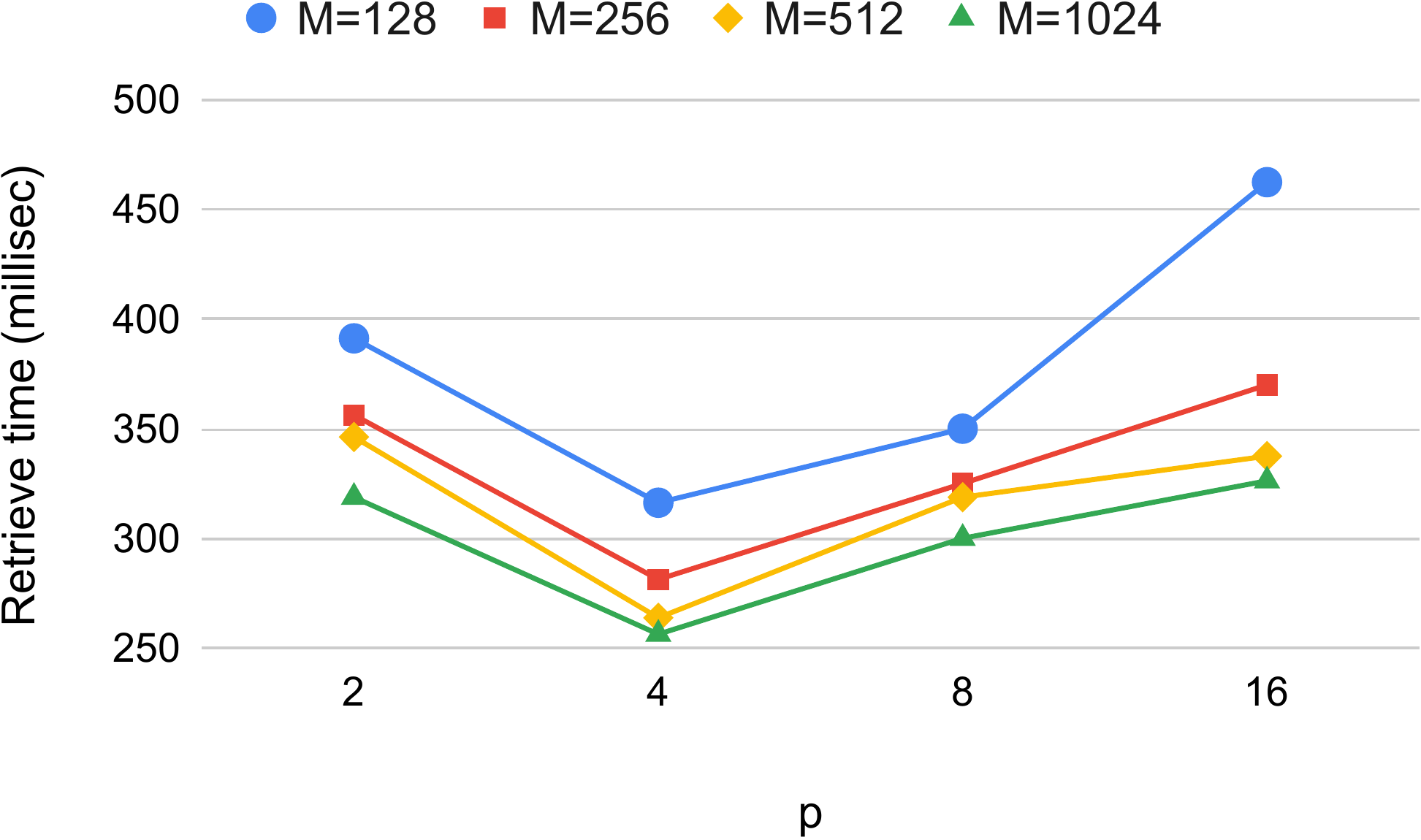}
        \caption{PP-KD-tree:Traject-1M}
        \label{fig:trajectory-1M_KD-tree_dim_recall}
    \end{subfigure}
    \caption{Sensitivity of ANN retrieval efficiency with respect to the dimensionality of projection ($p$) and the number of quantization intervals ($M$).}
    \label{fig:dim_recall}
\end{figure*}

\para{Parameters and Evaluation Metrics}

The two main parameters for privacy preserving data encoding are a) the dimension of the encoded space $p$ (Equation \ref{eq:transform}), and b) the number of equi-spaced intervals along each axis of the data, $M$, which is inversely proportional to the quantization interval $\delta$ (c.f. Equation \ref{eq:bounds} and \ref{eq:intervals}). 
For our experiments, we set $p$ to values of $2$, $4$, $8$, and $16$.
$M$ is chosen independently for each dataset depending on the density of the points.
For Traject-10K and Traject-100K datasets we set $M$ to $16$, $32$, $64$ and $128$, whereas for Traject-1M data we set it to $128$, $256$, $512$ and $1024$. Likewise for the FourSquare dataset (CheckIn-24M), we set its value to $1K$, $10K$, $100K$ and $1M$.


For the synthetic trajectory datasets, we conduct an ANN retrieval for each space-time coordinate of an infected (query) user. This means that a final list of susceptible candidates is obtained by aggregating (set union) of these individual lists. The size, say $r$, of the retrieved list at each distinct time coordinate value is set to 
Another parameter for the our ANN search is the number of retrieved suspected nearest neighbours, say $r$, of an infected person at each timestamp is varied from $10$ to $100$ in steps of $10$.

Since the task of finding susceptible candidates is a recall-oriented task (false negatives are less desirable), we evaluate the effectiveness of susceptible retrieval with recall, which measures the proportion of the true nearest neighbors (true susceptible candidates) that are eventually retrieved.


\subsection{Results}
Table \ref{tab:res} present the results of the different ANN search workflows on privacy-encoded data (named `PP-KD-tree' and `PP-HNSW'). The key observations from Table \ref{tab:res} are as follows. First, we observe that both the approaches yield satisfactory recall values which demonstrates the feasibility of applying an ANN-based workflow in pandemic situations to achieve a trade-off between recall and computation time. The time reported in milli-seconds refers to the time taken to retrieve a list for a single query. The retrieval times of both KD-tree and HNSW are substantially lower than an exhaustive search through the database (for the Traject-1M dataset the exhaustive search takes $17K\times$ more time on an average).

Although PP-KD-tree yields better recall values than PP-HNSW, the retrieval time of the KD-tree based ANN approach is higher than that of HNSW. It is seen that the recall values achieved are relatively insensitive to the number of people infected.

\textbf{Parameter Sensitivity.}
In Figure \ref{fig:retrieve_recall_time} we can see that by increasing the number of candidates retrieved per time step we can get better recall value but at the same time we can see that it increases retrieval time. We observed that setting \#retrieve/timestep as $100$ we can get near optimal result for all of the dataset within satisfactory retrieval time.

Figure \ref{fig:bin_recall} presents the sensitivity of PP-HNSW and PP-KD-tree with respect to the two parameters, namely the dimension of the basis vectors for projection ($p$) and the number of quantization grids ($M$).
From the figure, we observe that increasing the number of bins  increases recall values. However, we also note that it is not required to arbitrarily increase the value of $M$ because the results tend to saturate out with the use of $M=128$ bins for Traject-10/100K and for $M=1024$ in the case of Traject-1M dataset. Since the density of Check-in dataset is higher than those of the synthetic ones (see Table \ref{tab:dataset_summary}), the number of intervals required to achieve satisfactory recall values is also higher for this dataset. Figure \ref{fig:bin_recall} shows that about $1$M intervals in both PP-HNSW and PP-KD-tree are required to obtain satisfactory recall values.
%
Similar to investigating the recall variations, in Figure \ref{fig:dim_recall} showcases the effects of varying the parameters $p$ and $M$ on ANN retrieval time.


\section{Conclusions}   \label{sec:concl}
In this paper, we investigated the feasibility of applying standard approximate nearest neighbor (ANN) search approaches for the task of contact tracing in pandemic situations. More concretely, given an indexed collection of space-time coordinates of individuals and a list of infected persons, our intention is to retrieve a list of candidate persons might be susceptible to the infection since they came in close proximity (approximately same place and time) with the people already infected.
Since location data for contact tracing could lead to privacy issues, we also propose to an encoding and quantization based obfuscation of the data.
We conduct a set of laboratory-based experiments on data with known ground-truths. We found that the recall values that could be achieved with ANN-based approaches are satisfactory. Although the recall levels do decrease with an increase in the number of data points, our experiments show that for large datasets ANN based retrieval can achieve speed-gains of up to $17$K, thus achieving a relative trade-off between run-time and accuracy. These savings in run-time could be pivotal for early identification of susceptible cases and carry out necessary measures (e.g. quarantine the susceptible persons) for the health-care safety of a community. The proposed workflow also ensures that it is not required to share true user locations for contact tracing purposes. Instead, such a methodology for contact tracing in pandemic situations works fairly well with distance-preserving transformation of the data.


\bibliographystyle{plainnat}
\bibliography{main_arxiv}


\end{document}